\documentclass[12pt]{iopart}

\usepackage{iopams}
\usepackage{bm}
\usepackage{graphicx}
\usepackage{color}

\begin{document}

\newcommand{\beq}{\begin{equation}}\newcommand{\eeq}{\end{equation}}
\newcommand{\barr}{\begin{eqnarray}}\newcommand{\earr}{\end{eqnarray}}

\newcommand{\andy}[1]{ }

\newcommand{\ket}[1]{\left| #1 \right\rangle}
\newcommand{\bra}[1]{\left\langle #1 \right|}
\def\wtilde{\widetilde}
\def\Ord{\mbox{O}}
\def\As{{\cal A}}
\def\cH{{\cal H}}
\def\cN{{\cal N}}
\def\cP{{\cal P}}
\def\cR{{\cal R}}
\def\cV{{\cal V}}
\def\cU{{\cal U}}
\def\cZ{{\cal Z}}
\def\chiA{\chi_{_{A}}}
\def\de{\mathrm{d}}
\renewcommand{\Re}{{\mathrm{Re}}\, }
\renewcommand{\Im}{{\mathrm{Im}}\, }

\newcommand{\rev}[1]{{\color{red}#1}}
\newcommand{\REV}[1]{\textbf{\color{red}#1}}
\newcommand{\BLUE}[1]{\textbf{\color{blue}#1}}
\newcommand{\GREEN}[1]{\textbf{\color{green}#1}}


\topical[Quantum Zeno dynamics]{Quantum Zeno
dynamics: mathematical and physical aspects}


\author{P Facchi$^{1,2}$ and S Pascazio$^{3,2}$}

\address{$^{1}$ Dipartimento di Matematica, Universit\`a di Bari,
        I-70125  Bari, Italy}
\address{$^{2}$ Istituto Nazionale di Fisica Nucleare, Sezione di Bari, I-70126 Bari, Italy}
\address{$^{3}$ Dipartimento di Fisica, Universit\`a di Bari,
        I-70126  Bari, Italy}
\eads{\mailto{paolo.facchi@ba.infn.it},
\mailto{saverio.pascazio@ba.infn.it}}

\begin{abstract}
If frequent measurements ascertain whether a quantum system is still
in its initial state, transitions to other states are hindered and
the quantum Zeno effect takes place. However, in its broader
formulation, the quantum Zeno effect does not necessarily freeze
everything. On the contrary, for frequent projections onto a
multidimensional subspace, the system can evolve away from its
initial state, although it remains in the subspace defined by the
measurement. The continuing time evolution within the projected
``quantum Zeno subspace" is called ``quantum Zeno dynamics:" for
instance, if the measurements ascertain whether a quantum particle
is in a given spatial region, the evolution is unitary and the
generator of the Zeno dynamics is the Hamiltonian with hard-wall
(Dirichlet) boundary conditions. We discuss the physical and
mathematical aspects of this evolution, highlighting the open
mathematical problems. We then analyze some alternative strategies
to obtain a Zeno dynamics and show that they are physically
equivalent.

\end{abstract}



\newpage

\tableofcontents

\newpage


\section{Introduction: philosophy, history and physics}
 \label{sec-introd}
 \andy{intro}

Zeno was a sophist philosopher, and a native of Elea, in Southern
Italy. The quantum Zeno effect is named after him
\cite{Misra77}. In this article we will discuss the physical and
mathematical aspects of this phenomenon and its dynamical features.

Zeno put forward some famous paradoxical arguments against motion.
We shall focus on one of these, according to which a sped arrow does
not move, and analyze under which conditions an evolving quantum
state may not move. We will start by framing Zeno's arguments and
their quantum mechanical counterparts in the proper philosophical,
historical, cultural and physical context. This is a technical
article, that will be (hopefully) read by physicists and
mathematicians, yet we deem it appropriate to start from a genuine
humanistic perspective.

\subsection{Historical and philosophical prelude}
 \label{sec-introdhist}
 \andy{introdhist}

Zeno was born about 488 BC in Elea, a small town not far from
Naples, in the Mediterranean region called ``Magna Graecia". He was
a prominent figure of the Eleatic school of philosophy, founded by
Parmenides, of whom he was the favorite disciple. They believed that
senses are deceptive and motion and change mere illusions: there is
only one {\em Truth} (``being") that is static, does not change and
cannot be decomposed into parts or smaller entities. It is therefore
indivisible and does not develop. This view contrasted with the
notion of reality of Pythagoras and Heraclitus, who believed in a
world of change and ``becoming" and started revealing the power of
thought and mathematics, on which modern science stands
\cite{russell1,russell,hardy}.

Zeno's arguments were subtle and profound and aimed at bringing to
light some paradoxical aspects of common sense and of the notion of
a reality in continuous change. Some of these arguments directly
deal with motion and one in particular will concern us here: a
flying arrow is at rest. Indeed, at any given moment the arrow is in
a portion of space equal to its own length, and therefore is at rest
at that moment. Therefore, at every instant the arrow is motionless
in a given position and the ``sum" of these positions of rest is not
a motion. Notice that, according to the original sources
\cite{aristotle}, no direct mention is made to the act of observation. Yet,
implicitly, Zeno argued against our sensorial (and therefore
deceptive) perception of movement.

\begin{figure}
\includegraphics[width=\textwidth]{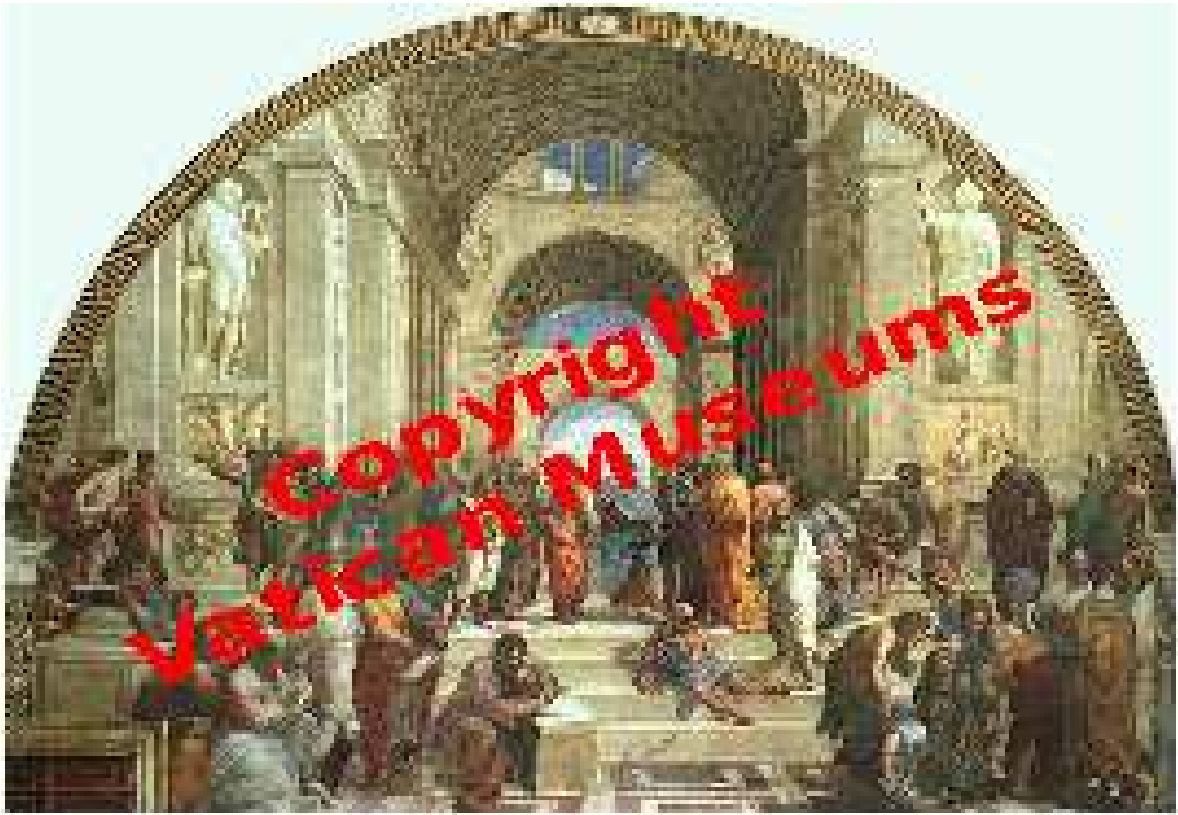}
\caption{The School of Athens was
painted by Raphael between 1509 and 1510 (in the frescoed Room of
the Signature, one of Raphael's rooms, in the Vatican Museums). All
the protagonists of our story appear in this fresco. Parmenides is
standing at the center-left, in a yellow shirt. Heraclitus
(portrayed with Michelangelo's features) is sitting at the center,
his hand holding his head. Pythagoras, dressed in white and red,
writes in a book, at the front-left. Zeno is at the very left, with
a white beard, partially hidden by a putto (the baby holding a
book): curiously, he looks older than his master Parmenides
(although, according to Plato
\cite{aristotle}, he was twenty-five years younger). Plato and
Aristotle are the dominant figures at the center of the painting and
Raphael himself is standing at the very right, looking at us; he is
almost hidden by Sodoma. (Courtesy Vatican Museums.)}
\label{fig:atene}
\end{figure}

Parmenides and Zeno went to Athens to discuss their ideas with Plato
and Aristotle. We learn from Plato that at the time of that trip
Zeno was about 40 years old, Parmenides about 65 and the pupil put
all his energy in defending the view of his master. However, the two
Eleatics were defeated (those were times when solidity of
argumentation made a difference in philosophical and political
discussions). Although many ancient writers refer to Zeno's ideas,
none of his writings survives, so that most of what we know of him
comes from what Aristotle wrote
\cite{aristotle}. This is like learning about a manuscript by
reading a (negative) referee report, although an authoritative one.
See Figure \ref{fig:atene}.

Plato calls Zeno the inventor of dialectic: his arguments foreran
the seemingly paradoxical method of proof that is known nowadays as
\emph{reductio ad absurdum}. It is therefore not surprising that
Zeno is considered a precursor of sophism. The implication of the
word sophist has changed greatly over time, running from its
original meaning. Initially, it was a highly complimentary term,
referred to someone who conveyed knowledge and wisdom (``sophia") to
his disciples. Eventually, by the time of Plato and Aristotle, the
word had taken on negative connotations, usually referring to
someone who used the arts of debate and rhetoric in order to
deceive, or to support fallacious reasoning. In some modern
Indo-European languages, words that have the same root as
``sophisticator" have acquired a very negative meaning. In the
Webster's dictionary one finds that ``echoes of Sophism survive
today in the language theory of Jacques Derrida and other postmodern
rhetoricians who teach that language ought to be deconstructed in
order to unpack the intentions of ``sophisticated" communicators."
Contemporary physics has sometimes, righteously, a difficult
relationship with postmodern philosophy
\cite{SokalBricmont,Weinberg}.

Yet this inherited negative attitude does not render justice to
Zeno, whose paradoxes have inspired philosophers for over two
millennia, challenged mathematicians for a few hundred years, and
puzzled physicists for the last 30 years. Many great thinkers looked
for a way out of the paradoxes. Kant, Hume, Hegel and many others
proposed solutions to Zeno's paradoxes, pondering over the meaning
of space, time and our forms of perception. Although these solutions
were ultimately not accepted by modern mathematics and science, they
helped shaping our modern concepts and rigorous definitions: our
ideas of space, time, motion, infinite, infinitesimal, line, point,
derivative and measure would not be the same without Zeno's input.
After two thousand years of continual refutation, his arguments made
the foundation of modern mathematics. Rigorous definitions are
unavoidable if one wants to rely on a logically consistent
mathematical scheme and does not want to fall into contradictions.

It is probably fairly accurate to say that Zeno's problems stem from
human efforts to comprehend the infinite. Nowadays it is common to
hear, in mathematics and physics circles, that Zeno's arguments are
based on false assumptions and that no Zeno-like paradoxes are
present within modern mathematics. We venture to disagree, not on
the conclusions, but on the premises: modern mathematics was created
in order to avoid those questions that Zeno posed, in exactly the
same way as modern set theory carefully avoids the use of Russell's
set of sets. Sped arrows move, runners catch tortoises, one can pass
an infinite number of points in a finite time and our concepts of
infinitesimal, finite and infinite have been adapted to our need to
arrive at a consistent description of motion \cite{xoff}. Clearly,
no subsequent logically consistent system could afford to ignore
Zeno's paradoxes, but of course it would have been impracticable to
accept passively a doctrine like that of Elea. When we describe the
motion of a ball in terms of Newton's law or the evolution of a wave
function in terms of Schr\"{o}dinger's equation, we stand on the
shoulders of thinkers who solved seemingly irreconcilable
contradictions.

\subsection{The quantum Zeno effect}
 \label{sec-introphys}
 \andy{introphys}

The afore-mentioned Zeno's paradox will be the object of the present
investigation: A sped arrow will never reach its target, because
when we look at the arrow, we see that it occupies a portion of
space equal to its own size. At any given moment the arrow is
therefore immobile, and by summing up many such ``immobilities" it
is impossible, according to Zeno, to obtain motion. It is amusing
that quantum mechanical systems, when observed, behave in a way that
is reminiscent of this paradox.

Frequent measurement slow the evolution of a quantum system,
hindering transitions to states different from the initial one. This
phenomenon, known as the \emph{quantum Zeno effect} (QZE)
\cite{Misra77,vonNeumann32,Beskow67,Khalfin68}, is a consequence of
general features of the Schr\"odinger equation, that yield quadratic
behavior of the survival probability at short times
\cite{rev1,rev2,rev3}.

Von Neumann \cite{vonNeumann32} was the first to understand the
far-reaching outcomes of the short-time evolution and bring to light
the essentials of the Zeno phenomenon. However, he was focusing on
quantum thermodynamics and his conclusions were forgotten for 35
years, when Beskow and Nilsson in a brilliant article
\cite{Beskow67} argued that frequent position measurements,
implicitly carried out by a bubble chamber on an unstable particle,
could prevent its decay. These ideas were  corroborated by Khalfin
\cite{Khalfin68}, 10 years after his seminal work on the long-time
features  of quantum evolutions
\cite{Khalfin57,Khalfin58}, and were finally put on firm mathematical
ground by Friedman \cite{Friedman72}.

At the same time, there was a widespread mathematical interest on
the convergence of the Trotter-Kato product formula
\cite{Trotter1,Trotter2,Kato1,HillePhillips}.
Mathematicians  focused on the limit of product formulas when the
potential is singular or the Hamiltonian unbounded
\cite{Chernoff,gustafsonmisra,Friedman1,gustafson,Exner,gustafson2,Schmidt02,Schmidt03,Schmidt04,
Matolcsi03,ExnerIchinose,EINZ}.
These limits and their properties are indeed of great
importance in the  study of quantum dynamical semigroups and may
contain the germs of irreversibility. Their relevance is therefore
twofold, as they have remarkable consequences both in mathematical
physics and operator theory.

Finally, these ideas were
gathered by Baidyanaith Misra and George Sudarshan in their
pioneering article
\cite{Misra77}, that introduced, among other things, the classical
allusion to the Eleatic philosopher. The paper by Misra ad Sudarshan
is also amusing in its own right: it blends rigorous mathematics
with subtle and often ironical remarks about philosophy and cats.
Since then, the QZE received constant attention by theorists, who
explored different facets of the phenomenon.

The Zeno problem was considered an academic one until 1988, when
Cook \cite{Cook88} proposed a test with oscillating systems, rather
than on \textit{bona fide} unstable ones. This was a concrete idea,
that revived the subject and led to the celebrated experimental test
by Itano \emph{et al} a few years later \cite{Itano90}. The
discussion that followed
\cite{Petrosky90,PeresRon,Itano91,Petrosky91,Inagaki,Frerichs92,Pascazio93,%
altenmuller,Pascazio94,Schulman94,Berry95,beige,KK96,Schulman97,%
Mihokova97,lss98,LuisPerina,Facchi99b,HW2,whit} provided insight and new
ideas, eventually leading to new experimental tests. The QZE was
successfully checked in a variety of different situations, on
experiments involving photon polarization
\cite{kwiat}, nuclear spin isomers \cite{Chapovsky}, individual ions
\cite{Balzer,Toschek,Wunderlich,balzer2002}, optical pumping
\cite{molhave2000}, NMR \cite{jones}, Bose-Einstein condensates
\cite{ketterle} and new experiments are in preparation with neutron spin
\cite{VESTA,RauchVESTA} and superconducting qubits \cite{Johansson,vion}.
One should also emphasize that the first experiments were not free
from interpretational problems. Some of these were successfully
solved (e.g., the issue of the so-called ``repopulation" of the
initial state \cite{Nakazato96b} was first avoided in
\cite{Balzer}), but some authors (sensibly) argued that the QZE had
not been successfully demonstrated on \emph{bona fide} unstable
systems, as in the seminal proposals. Indeed, all the above
mentioned experiments deal with finite (oscillating) systems, whose
Poincar\'e time is finite. The analysis of the evolution at short
times becomes more complicated when one considers unstable systems
\cite{Rimini,Bernardini93,Facchi98,Maiani98,Joichi98,Facchi99c,Alvarez99}: in such a
case, novel and somewhat unexpected phenomena come to light. There
were also noteworthy early attempts at interpreting preexisting
experimental data on unstable particle decay \cite{valanju}.
However, only two experiments have been successfully performed so
far on unstable systems, both by Mark Raizen's group. In the first
one
\cite{Wilkinson97} the presence of the short-time non-exponential
region was brought to light, while in the second one
\cite{raizenlatest} the quantum Zeno effect and its inverse were
demonstrated. It is worth emphasizing that recent
 experimental demonstrations of the QZE are driven by interest
in fundamental physics, but also by practical applications, such as
efficient preservation of spin polarization in gases
\cite{Chapovsky,nakanishi2001}, dosage reduction
in neutron tomography \cite{tomography2002} and control of
decoherence in quantum computing
\cite{franson2004,FTPNTL2005,sasaki2005,hosten2006}.
The last ten years have witnessed a number of very
interesting ideas on the quantum Zeno effect as well as on possible
applications
\cite{Blanchard93,cirac,Zhu95,beige,Huang96,Zhu96,Facchi99b,Facchi00b,MMN,Militello01,KK,pastawski,distant,maniscalco,longhi,tian,gordon,maniscalco2}.
Some recent proposals, aimed at countering the detrimental effects
of decoherence, deal with closely related quantum dynamics, in which
the quantum measurements are frequent but not infinitely frequent
\cite{NTY,NUY,WLS,compagno,MM,YN,paternostro,MYNM,NYMM}.
The whole field is very active.

\subsection{Quantum Zeno dynamics and quantum Zeno subspaces}
 \label{sec-multifin}
 \andy{multifin}

Although all the afore-mentioned experiments invigorated studies on
this issue, they essentially  deal with one-dimensional projections
(and therefore with what we might call  one-dimensional Zeno
effect): the system is forced to remain in its initial state.
However, the QZE does not necessarily freeze everything. On the
contrary, for a projection onto a \emph{multidimensional} subspace,
the system can evolve away from its initial state, although it
remains in the subspace defined by the ``measurement." This
continuing time evolution {\em within} the projected ``Zeno
subspace" we call {\em quantum Zeno dynamics} and is the central
topic of this paper. It is often overlooked, although it is readily
understandable in terms of the seminal theorem by Misra and
Sudarshan \cite{Misra77}. One PACS number is now devoted to the
quantum Zeno dynamics. Interestingly, less than one decade ago, an
anonymous referee asked that the expression ``quantum Zeno dynamics"
be removed from the title of one of our articles. Times
change, PACS numbers too.

We shall also discuss an interesting idea that has been repeatedly
put forward in the mathematical and physical literature: Zeno
dynamics yields ordinary constraints. In particular, suppose a
system has Hamiltonian $H$ and the (frequent) measurement is
checking that the system is within a particular spatial region. Then
the Zeno dynamics that results is governed by the same Hamiltonian,
but with Dirichlet boundary conditions on the boundary of the
spatial region associated with the projection.

No experiments have been performed so far in order to check the
multidimensional Zeno effect and the Zeno dynamics. However, these
ideas might lead to remarkable applications, e.g.\ in the control of
decoherence. For this reason, we shall explicitly look at some
simple finite-dimensional examples of Zeno dynamics, in order to
show how to freeze the dynamics of the simplest multidimensional
system: a qubit.

\subsection{Outline}
 \label{sec-outline}
 \andy{outline}

This review article is organized as follows. We take a physicist's
approach in Secs.\
\ref{sec-QZE2lev}-\ref{sec-infinitedim},
where the QZE is introduced and discussed without spelling out
implicit hypotheses and with no rigor in the derivations. The
mathematics behind the QZE is discussed in Secs.\
\ref{sec-bounded}-\ref{sec-coroll}. We focus here on the state of the art and explain what
are the main mathematical difficulties and open problems. We finally
turn to the practical applications of the QZE in Secs.\
\ref{sec-zenosubsp}-\ref{sec-example}: we show that it can be used
in order to control the quantum dynamics and give three alternative
(but physically equivalent) derivations of the Zeno subspaces. We
conclude with an outlook in Sec.\
\ref{sec-final}.

The path we will choose to describe the QZE will depend on our taste,
for both its physical and mathematical aspects. A physically
oriented reader can skip Secs.\
\ref{sec-bounded}-\ref{sec-coroll}. A mathematically oriented reader can instead skip
Secs.\ \ref{sec-zenosubsp}-\ref{sec-example}. However, we believe
that a thorough comprehension of the Zeno phenomena, its
implications and potential applications, can only be gained by
combining physical and mathematical insight.

\section{One-dimensional case}
\label{sec-QZE2lev}
\andy{QZE2lev}

\subsection{Fundamentals}
\label{sec-QZEpuls}
\andy{QZEpuls}

Let a quantum system be prepared, at time $t=0$, in the pure state
$|\psi_0\rangle$, a normalized vector in the Hilbert space $\cH$.
The system evolves under the action of the total Hamiltonian $H$.
The quantities
\andy{survampl}
\barr
\As(t) & = & \langle \psi_0|\psi_t\rangle = \langle
\psi_0|\mathrm{e}^{-iHt}|\psi_0\rangle, \\
\label{eq:survampl}
p(t) & = & |\As (t)|^2 =|\langle \psi_0|\mathrm{e}^{-iHt}|\psi_0\rangle |^2 ,
\label{eq:survpr}
\earr
are called {\em survival amplitude} and {\em probability},
respectively, and represent the amplitude and probability that the
quantum system is found in the initial state $|\psi_0\rangle$ at
time $t$. An elementary expansion at short times yields a quadratic
behavior
\beq
p(t) = 1 - t^2/\tau_{\rm Z}^2 + \ldots,
\qquad
\tau_{\rm Z}^{-2} \equiv \langle\psi_0|H^2|\psi_0\rangle -
\langle\psi_0|H|\psi_0\rangle^2 ,
\label{eq:naiedef}
\eeq
where $\tau_{\rm Z}$ is the ``Zeno time," a (sometimes very
inaccurate) quantitative estimate of the duration of the short-time
quadratic behavior. Notice that if one decomposes the total
Hamiltonian into a free and an interaction part
\barr
H&=&H_0 + H_{\rm int},\nonumber\\
H_0 &=&P H P + Q H Q, \qquad H_{\rm int}=P H Q + Q H P,
\label{eq:decomp}
\earr
where $P=\ket{\psi_0}\bra{\psi_0}$ and $Q=1-P$, the initial state is
an eigenstate of the free Hamiltonian, that is also completely
off-diagonal with respect to the interaction
\beq
\label{eq:decomp11}
H_0\ket{\psi_0}=\omega_0\ket{\psi_0},\qquad \bra{\psi_0}H_{\rm
int}\ket{\psi_0}=0.
\eeq
The Zeno time reads then
\beq
\tau_{\rm Z}^{-2} = \langle\psi_0|H_{\rm int}^2|\psi_0\rangle
\eeq
and depends only on the square of the interaction Hamiltonian. While
the Fermi golden rule
\cite{Fermi32,Fermi50,Fermi54} ``picks" only the on-shell contribution
of the interaction, the Zeno time ``explores" (in agreement with the
time-energy uncertainty relation) all possible intermediate states,
by virtue of the completeness relation
\beq
\label{eq:zenoexplores}
\langle\psi_0|H_{\rm int}^2|\psi_0\rangle =
\langle\psi_0|H_{\rm int} Q H_{\rm int}|\psi_0\rangle =
\sum_{n\neq 0}
\langle\psi_0|H_{\rm int}|n\rangle\langle n|H_{\rm int}|\psi_0\rangle,
\eeq
where $|n\rangle$ (with $\ket{0}=\ket{\psi_0}$) is an eigenbasis of
the free Hamiltonian $H_0$. The decomposition (\ref{eq:decomp}) is
at the heart of the Lippmann-Schwinger equation \cite{Sakurai} and
plays an important role in quantum field theory.

\begin{figure}
\includegraphics[width=.6\textwidth]{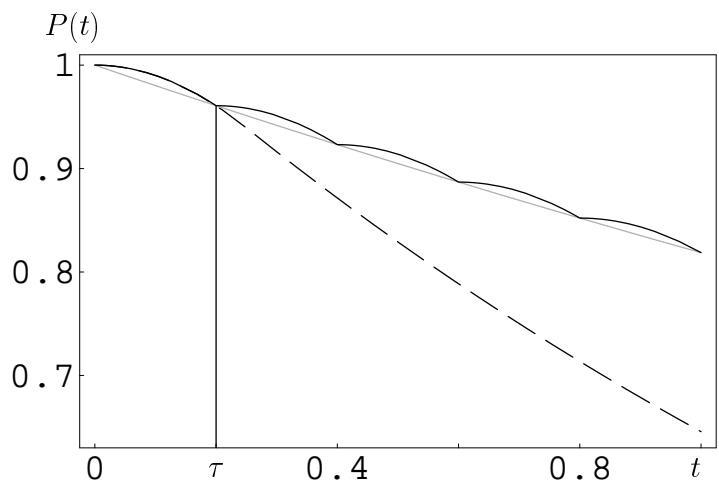}
\caption{The quantum
Zeno effect. The dashed (full) line is the survival probability
without (with) measurements. The total duration of the experiment is
$t=1$ and there are $N=5$ measurements, so that $\tau=0.2$. In the
limit of ``continuous observation" ($N \to \infty$, $t$ finite),
$p(t) \to 1$. The grey line is the interpolating function in the far
r.h.s.\ of Eq.\ (\ref{eq:survN0}).}
\label{fig:zenoevol}
\end{figure}

Let us now carry out $N$ measurements at time intervals $\tau=t/N$,
in order to check whether the system is still in its initial state.
If every time the measurement has a positive outcome and the system
is found in its initial state, the wave function ``collapses" and
the evolution starts anew from $|\psi_0\rangle$. The survival
probability after the $N$ measurements reads
\andy{survN}
\barr
p^{(N)}(t)&=&p(\tau)^N = p(t/N)^N\nonumber\\
&\stackrel{N \; {\rm large}}{\sim}& \left[ 1 - (t/N\tau_{\rm Z})^2
\right]^N \sim \exp(-t^2/N\tau_{\rm Z}^2)
\stackrel{N \rightarrow\infty}{\longrightarrow} 1 ,
 \label{eq:survN}
\earr
where $t=N\tau (<\infty)$ is the total duration of the experiment.
The Zeno evolution is pictorially represented in Fig.\
\ref{fig:zenoevol}. The $N \rightarrow\infty$ limit was originally
named limit of ``continuous observation"  by Misra and Sudarshan and
regarded as a paradoxical results: infinitely frequent measurements
halt the quantum mechanical evolution and freeze the system in its
initial state. Zeno's quantum mechanical arrow (the wave function),
sped by the Hamiltonian, does not move, if it is continuously
observed. The QZE is not considered paradoxical nowadays: it is a
consequence (admittedly, a curious one) of the quantum mechanical
evolution law. Clearly, if one considers real physical measurement
processes, effectively described in terms of projection operators,
the $N \rightarrow \infty$ limit must  be considered as a convenient
mathematical abstraction
\cite{Nakazato95,Venugopalan95,Pati96,Hradil98}, but the evolution
is indeed slowed down for sufficiently large $N$. An interesting
alternative viewpoint can be found in
\cite{Berry96}, where the ``continuous" measurement is implemented
in terms of a dynamical constraint. From such a perspective, no $N
\to \infty$ limit need be considered. The constraint picture that
arises from the QZE is discussed in \cite{FMPSS}.

As a very simple example consider a two-level system with
Hamiltonian
\andy{hamrabi1}
\beq
H=H_{\rm int} = \Omega \sigma_1 ,
\label{hamrabi}
\eeq
where $\Omega\in \mathbb{R}$ and $\sigma_1$ is the first Pauli
matrix. Let the initial state be $|\psi_0\rangle = |+\rangle$, the
positive eigenstate of the third Pauli matrix $\sigma_3$. The
survival amplitude, survival probability and Zeno time read
\andy{tauzrabi}
\beq
\As (t) = \langle + | \mathrm{e}^{-i \Omega t \sigma_1} |+\rangle =
 \cos \Omega t, \qquad p(t)  =  \cos^2 \Omega t, \qquad
\tau_{\rm Z}  =  \Omega^{-1},
\label{eq:tauzrabi}
\eeq
respectively. For large $N$,
\beq
p^{(N)}(t) = \left( \cos \frac{\Omega t}{N} \right)^{2N} \sim
\left( 1 - \frac{\Omega^2 t^2}{2N^2} \right)^{2N}
\sim \mathrm{e}^{-\Omega^2t^2/N} \stackrel{N \to \infty}{\longrightarrow}
1 .
\label{eq:Nlargeex}
\eeq
The limit (\ref{eq:survN}) is therefore valid and one obtains a nice
example of QZE. Observe that in this case $\tau_{\rm Z}$ does yield
a good estimate of the duration of the short time region.

\subsection{A few preliminary comments and the inverse Zeno effect}
\label{sec-QZEinv}

Let us briefly comment on the seemingly innocuous derivation of the
preceding subsection.

\subsubsection{Short-time dynamics.}
\begin{figure}
\includegraphics[width=.6\textwidth]{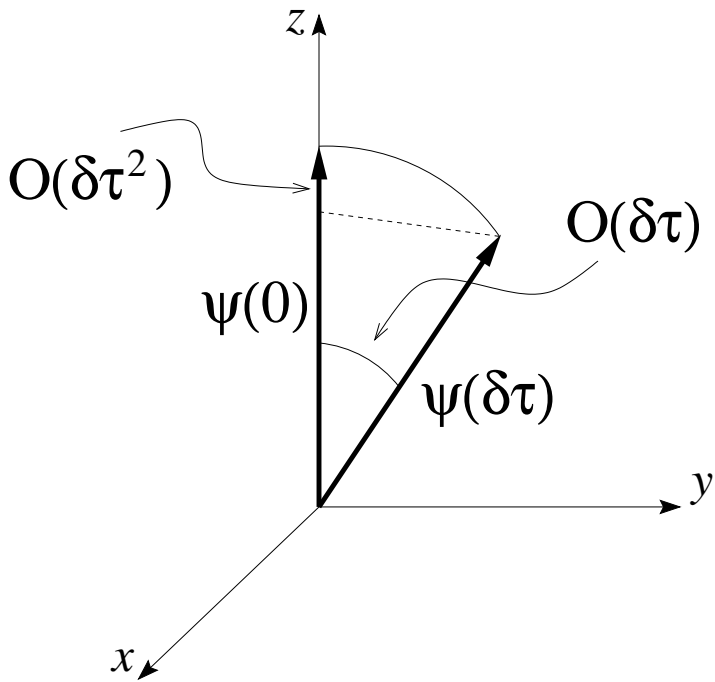}
\caption{Short-time evolution of phase and probability.
The phase of the evolved wave function is linear in time, while the
probability loss is quadratic. }
\label{fig:phprob}
\end{figure}

The QZE is ascribable to the following mathematical feature of the
Schr\"odinger equation: in a short time $\delta\tau \propto 1/N $, the
phase of the wave function evolves like $\Ord (\delta\tau)$, while
the probability changes by $\Ord (\delta\tau^2)$, so that
\andy{survN3}
\beq
p^{(N)}(t) = \left[ 1 - \Ord (1/N^2)\right]^N \stackrel{N
\rightarrow \infty}{\longrightarrow}1.
 \label{eq:survN3}
\eeq
This is sketched in Fig.\ \ref{fig:phprob}.

\subsubsection{Quantum measurements.}
In the preceding analysis, the measurements simply ascertain whether
the system is in its initial pure state. In other words, the von
Neumann measurement can be described by the
\emph{one-dimensional} projector
\andy{proj0}
\beq
P= |\psi_0\rangle \langle\psi_0| .
 \label{eq:proj0}
\eeq
This notion will be extended in the following section, where more
general measurements will be considered. This will lead us to the
notion of Zeno subspaces.

\subsubsection{Rigor.} No care was taken of
subtle mathematical issues, such as for instance the validity of the
asymptotic expansion (\ref{eq:naiedef}). Loosely speaking, we
required finite moments of $H$ in the initial state and implicitly
assumed that the Hamiltonian $H$ be bounded. In a general setting,
the features of the Hamiltonian must be spelled out and it is not
obvious that the series (\ref{eq:naiedef}) will converge, or it is
even defined, as it involves higher order moments of the field
Hamiltonian in the cumulant expansion. An explicit example involving
the hydrogen atom can be found in \cite{Facchi98}. See also
\cite{Seke94a,Seke94b}, were the role of the form factors of the
interaction is scrutinized.

The quantum Zeno effect in a quantum field theoretical framework has
been investigated only relatively recently
\cite{Bernardini93,Facchi98,Maiani98,Joichi98,Facchi99c,Alvarez99,Antoniou}.
The features of the short-time evolution are not obvious, in
particular when the convergence of the expansions cannot be proven.

There are other problems that are largely unexplored. Among these,
the conditions that lead to a non-quadratic evolution at short times
\cite{Muga,anom0,anom}. Interestingly, these issues are relevant in quantum tunneling
\cite{arrival}.

\subsubsection{The Zeno time.} In many situations, the Zeno time
(\ref{eq:naiedef}) yields a misleading estimate: strictly speaking,
$\tau_{\rm Z}$ is nothing but the convexity of $p(t)$ in the origin.
If the asymptotic series (\ref{eq:naiedef}) is nasty, one needs a
much more refined estimate of the duration of the short-time
quadratic region \cite{Facchi98,Antoniou}. This leads us to the
final comment:

\subsubsection{Inverse Zeno effect: Heraclitus vs Zeno. }

Interfering with a transition at a later stage in its progress leads
to the opposite phenomenon, known as the inverse or anti-Zeno
effect, in which decay is accelerated. Both effects have recently
been seen in the same experimental setup \cite{raizenlatest}. Let us
briefly describe this interesting aspects of quantal evolutions.
Rewrite (\ref{eq:survN}) as
\andy{survN0}
\beq
p^{(N)}(t)=p(\tau)^N=\exp(N\log p(\tau))= \exp(-\gamma_{\rm
eff}(\tau) t) ,
 \label{eq:survN0}
\eeq
where we introduced the effective decay rate
\andy{effgamma0}
\beq
\gamma_{\rm eff}(\tau) \equiv -\frac{1}{\tau}\log p(\tau).
 \label{eq:effgamma0}
\eeq
The far right hand side of (\ref{eq:survN0}) interpolates the
survival probability when $N$ measurements are performed, namely it
intercepts the survival probability at its cusps, when the system
is projected back onto its initial state (see Fig.\
\ref{fig:zenoevol}).

Observe that in the region of validity of the expansion
(\ref{eq:naiedef}) $\gamma_{\rm eff}$ is a linear function of $\tau$
\andy{effgamma}
\beq
\gamma_{\rm eff}(\tau) \sim \tau/\tau_{\rm Z}^2, \qquad
\mbox{for}\quad \tau\to 0.
 \label{eq:effgamma}
\eeq
For example, with the Hamiltonian (\ref{hamrabi}) and the initial
state $|+\rangle$, one gets from Eq.\ (\ref{eq:Nlargeex})
\beq
p^{(N)}(t) \sim \mathrm{e}^{-\Omega^2t^2/N} = \mathrm{e}^{-(\Omega^2 \tau) t} =
\mathrm{e}^{-\gamma_{\rm eff}(\tau) t},
\label{eq:Nlargeexeff}
\eeq
with $\tau = t/N$ and $\gamma_{\rm eff}(\tau)\sim \Omega^2 \tau =
\tau/\tau_{\rm Z}^2$.

If the system is unstable one expects to recover the ``natural"
decay rate $\gamma$, in agreement with the Fermi golden rule, for
sufficiently long times, i.e., after the initial quadratic region is
over
\andy{effgammalim}
\beq
\gamma_{\rm eff}(\tau) \stackrel{{\rm ``large"} \tau}{\longrightarrow} \gamma.
 \label{eq:effgammalim}
\eeq
The physical meaning of the mathematical expressions $\tau \to 0$
and ``large" $\tau$ in the preceding formulas is in itself an
interesting issue, that entails the definition of a timescale
\cite{heraclitus,Facchi99c}. This difficult and delicate problem will not be
discussed here. Suffices it to say that when a quantum field
theoretical description is necessary, $\tau_{\rm Z}$ is not the
right timescale. We shall neither scrutinize here the features of
the quantum mechanical evolution law at short
\cite{Beskow67,Khalfin68} and long
\cite{Mandelstam45,Fock47,Hellund53,Namiki53,Khalfin57,Khalfin58}
times, nor analyze the validity of the Weisskopf-Wigner
approximation \cite{Gamow28,Weisskopf30a,Weisskopf30b,Breit36} and
the onset to the Fermi ``golden rule"
\cite{Fermi32,Fermi50,Fermi54}.
These topics are summarized and discussed in \cite{rev1}.

\begin{figure}
\includegraphics[width=\textwidth]{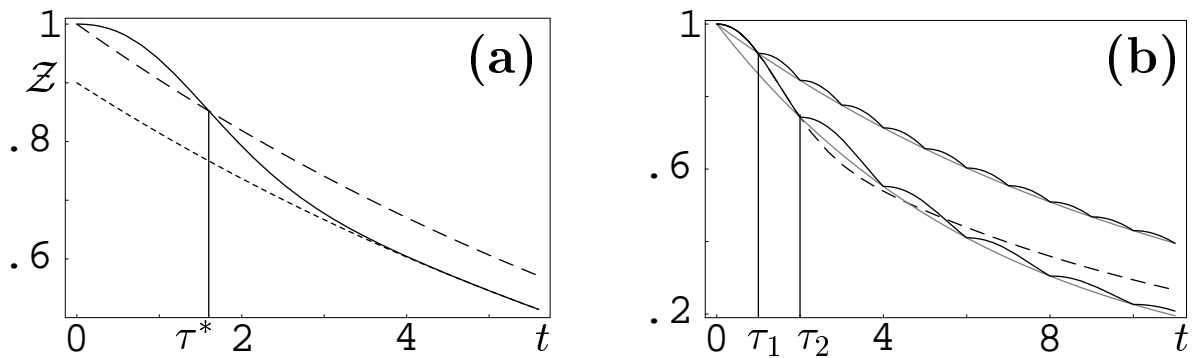}
\caption{(a) Time $\tau^*$ separates the quantum Zeno from the Heraclitus
(or inverse Zeno) regime. The full line is the survival probability
$p(t)$, that tends to the asymptotic exponential $\cZ \mathrm{e}^{-\gamma t}$
for large $t$ (dotted line, $\cZ$ stemming from wave-function
renormalization, see \cite{heraclitus}); the dashed line is the
exponential $\mathrm{e}^{-\gamma t}$. (b) Quantum Zeno vs inverse Zeno (or
Heraclitus) effect. The dashed line is the undisturbed survival
probability $p(t)$. The full black lines represent the survival
probabilities when measurements are made at time intervals $\tau$
and the dotted grey lines their exponential interpolations,
according to (\ref{eq:survN0}). For $\tau_1<\tau^*<\tau_2$ the
effective decay rate $\gamma_{\rm eff}(\tau_1)$ [$\gamma_{\rm
eff}(\tau_2)$] is smaller [larger] than the ``natural" decay rate
$\gamma=\gamma_{\rm eff}(\tau^*)$, yielding QZE [IZE].}
\label{fig:gtau}
\end{figure}

Consider now an unstable system, with decay rate $\gamma$. If a
finite time $\tau^*$ exists such that
\andy{tstardef}
\beq
\gamma_{\rm eff}(\tau^*)=\gamma,
\label{eq:tstardef}
\eeq
then by performing measurements at time intervals $\tau^*$ the
system decays according to its ``natural" lifetime, as if no
measurements were performed. By  Eqs.\ (\ref{eq:tstardef}) and
(\ref{eq:effgamma0}) one gets
\beq
p(\tau^*)=\mathrm{e}^{-\gamma \tau^*},
\eeq
i.e., $\tau^*$ is the intersection between the curves $p(t)$ and
$\mathrm{e}^{-\gamma t}$ \cite{heraclitus}. Figure\ \ref{fig:gtau}(a)
illustrates an example in which such a time $\tau^*$ exists. By
looking at Fig. \ref{fig:gtau}(b), it is evident that if
$\tau=\tau_1<\tau^*$ one obtains a QZE. {\em Vice versa}, if
$\tau=\tau_2>\tau^*$, one obtains an {\em inverse} Zeno effect
(IZE). In this sense, $\tau^*$ can be viewed as a {\em transition
time} from a quantum Zeno to an inverse Zeno regime. Paraphrasing
Misra and Sudarshan, we can say that $\tau^*$ determines the
transition from Zeno (who argued that a sped arrow does not move) to
Heraclitus (who replied that everything flows).

The Zeno-Heraclitus transition and the onset to the inverse Zeno
effect were discussed by different authors
\cite{lane,schieve,Pascazio96,KaulGontis,Schulman97,Thun98,Pascazio99,Facchi99a,Kofman99,Facchi00,kurizki,heraclitus}.
In some cases, $\tau^*$ does not exist and no inverse Zeno effect
can take place \cite{heraclitus}. In the remaining part of this
article we will assume that $N$ is sufficiently large, so that we
are in the Zeno regime, and that the $N \to \infty$ limit can be
taken.

\section{Finite-dimensional projections: the quantum Zeno subspaces}
\label{sec-finitedim}

We now give a broader definition of measurement and generalize the
notion of QZE. This can be done with the help of L\"uders's
postulate \cite{Luders}, that suitably extends von Neumann's
\cite{vonNeumann32}.

A measurement is called ``incomplete" if some outcomes are lumped
together, for instance because the measuring apparatus has
insufficient resolution. The projection operator that selects a
given lump is therefore multidimensional and in this sense the
information gained on the measured observable is incomplete. By
contrast, a (selective) complete measurement yields a definite
outcome of the observable being measured. In the discussion of Sec.\
\ref{sec-QZE2lev} the measurements were selective and complete, because the
system was found in $|\psi_0\rangle$.

Let the evolution of the quantum system in the Hilbert space $\cH$
be governed by the unitary operator $U(t)=\exp(-iHt)$, where $H$ is
a time-independent Hamiltonian. We assume that the projection
operator $P$ that describes the measurement does not commute with
the Hamiltonian, $[P,H]\neq 0$ and that $\mathrm{Tr} P=s<\infty$.
The measurement therefore ascertains whether the system is in the
$s$-dimensional subspace $P{\cal H}={\cal H}_P$. For instance, the
reader can think of a finite, say $m$-dimensional Hilbert space. In
such a case $H$ is a $m\times m$ matrix, with $\dim \cH_P = s<m
=\dim \cH$.

It is convenient to discuss the evolution in terms of density
matrices. The initial density matrix $\rho_0$ is taken to belong to
${\cal H}_P$ (state preparation)
\andy{inprep}
\beq
\rho_0 = P \rho_0 P , \qquad \mbox{Tr} [ \rho_0 P ] = 1
\label{eq:inprep}
\eeq
and the state at time $\tau$ is
\andy{noproie}
\beq
\rho (\tau) = U(\tau) \rho_0 U^\dagger (\tau)
  \label{eq:noproie}
\eeq
If we measure $P$ and the outcome is positive, the state, up to a
normalization $p(\tau)$, changes into
\cite{Luders}
\andy{proie}
\barr
\rho (\tau) & \rightarrow & P \rho(\tau) P=P U(\tau) \rho_0 U^\dagger(\tau) P
\nonumber \\
 &=& V(\tau) \rho_0 V^\dagger(\tau) ,
  \quad \qquad V(\tau) \equiv P U(\tau)P
\label{eq:proie}
\earr
and the survival probability in ${\cal H}_P$ reads
\andy{stillun}
\barr
p(\tau) & =& \mbox{Tr} \left[ U(\tau) \rho_0 U^\dagger(\tau) P
\right] = \mbox{Tr} \left[P U(\tau)P \rho_0 P U^\dagger(\tau) P \right]
\nonumber
\\ &=&
 \mbox{Tr} \left[V(\tau) \rho_0 V^\dagger(\tau) \right] .
\label{eq:probini}
\earr
Since $[P,H]\neq 0$, the Hamiltonian $H$ induces transitions out of
${\cal H}_P$ into ${\cal H}_P^\perp={\cal H}_{Q}$ $({\cal H}_P
\oplus{\cal H}_{Q} = \cH$ and $P+Q=1$)
and $p(\tau)$ is in general smaller than unity. There is, of course,
a probability $q(\tau)=1-p(\tau)$ that the system has not survived (i.e., it has
made a transition out of ${\cal H}_P$) and its state has changed, up to a normalization $q(\tau)$,
into
\barr
\rho (\tau) & \rightarrow &  Q \rho(\tau) Q= Q U(t) \rho_0 U^\dagger(t)
Q \nonumber \\
& = & V_{QP}(\tau) \rho_0 V_{QP}^\dagger(\tau) ,
  \quad \qquad V_{QP}(\tau) \equiv Q U(\tau)P.
\label{eq:proieperp}
\earr
The final state after the measurement is therefore a block diagonal
matrix:
\barr
\rho_0 & \stackrel{{\rm unitary}}{\longrightarrow} & U(\tau) \rho_0 U^\dagger(\tau)
\stackrel{{\rm measurement}}{\longrightarrow}
\pmatrix{V(\tau) \rho_0 V^\dagger(\tau) & 0 \cr 0 & V_{QP}(\tau)
\rho_0 V_{QP}^\dagger(\tau) }. \nonumber \\
\label{eq:blockdiag}
\earr
The density matrix is reduced to a mixture and any possibility of
interference between ``survived" and ``not survived" states is
destroyed (complete decoherence).

We shall henceforth concentrate our attention \emph{only} on the
measurement outcome (\ref{eq:proie})-(\ref{eq:probini}) and turn to
the multidimensional Zeno effect. The state of the system after a
(successful) series of $P$-observations at time intervals $\tau=t/N$
is
\andy{Nproie}
\beq
\rho^{(N)}(t) = V_N(t) \rho_0 V_N^\dagger(t)/p^{(N)}(t) , \qquad
    V_N(t) \equiv [ P U(t/N) P ]^N
\label{eq:Nproie}
\eeq
and the survival probability to find the system in ${\cal H}_P$ is
\andy{probNob}
\beq
p^{(N)}(t) = \mbox{Tr} \left[ V_N(t) \rho_0 V_N^\dagger(t)
\right].
\label{eq:probNob}
\eeq
We have to study the limit
\beq
U_{\rm Z} (t) \equiv
\lim_{N \to \infty} V_N(t) = \lim_{N \to \infty}\left[ P \mathrm{e}^{iHt/N} P \right]^N .
\label{eq:limnfin}
\eeq
This is easily computed by expanding
\begin{eqnarray}
V_N(t) &=& \left[P \left(1-i H t/N +O\left(1/N^2\right)
\right)P \right]^N
\nonumber\\
&=& P \left[1-i P H P t/N + O\left(1/N^2\right)
\right]^N \nonumber\\
& \stackrel{N \to \infty}{\longrightarrow} & P \mathrm{e}^{-i P H P t} =
U_{\rm Z} (t).
\end{eqnarray}
The dynamics is governed by the ``Zeno" Hamiltonian $ H_{\rm Z} \equiv P H P
$ and the evolution is unitary in
$\cH_{P}$. We shall write
\beq
U_{\rm Z} (t)=P \exp(-i H_{\rm Z} t)  \label{eq:UZeno}
\eeq
and speak of \emph{quantum Zeno dynamics} in the \emph{quantum Zeno
subspace} ${\cal H}_P$. Notice that the one-dimensional result of
the previous section is obtained when $\mbox{Tr}P={\cal H}_P=s=1$
(and then $H_{\rm Z}=$ constant = phase). The final state is
\andy{infproie}
\beq
\rho (t)= \lim_{N \rightarrow \infty} \rho^{(N)}(t)
= U_{\rm Z} (t) \rho_0 U_{\rm Z} (t)^\dagger
  \label{eq:infproie}
\eeq
and the probability to find the system in ${\cal H}_P$ is
\andy{probinfob}
\beq
\lim_{N \rightarrow \infty} p^{(N)}(t) =\mbox{Tr} \left[ U_{\rm Z} (t) \rho_0 U_{\rm Z} (t)^\dagger
\right] = \mbox{Tr} \left[  \rho_0 P
\right]= 1.
\label{eq:probinfob}
\eeq
This is the multidimensional QZE. If the particle is constantly
checked for whether it has remained in ${\cal H}_P$, it never makes
a transition to ${\cal H}_P^\perp$.

A few comments are in order. First, notice that for finite $N$ the
dynamics (\ref{eq:Nproie})-(\ref{eq:probNob}) is
not reversible. The dynamics becomes unitary and
reversible in the $N \to \infty$ limit. The physical mechanism that
ensures the conservation of probabilities within the relevant
subspace hinges on the short time behavior of the survival
probability: probability leaks out of the subspace ${\cal H}_P$ like
$\tau^2$ for short times. The infinite-$N$ limit suppresses this loss.
Finally, the analysis of this section is straightforward for finite
systems and finite dimensional projectors. Things get much more
complicated for infinite dimensional systems. One can then inquire
under what circumstances $U_{\rm Z} (t)$ actually forms a group,
yielding reversible dynamics within the Zeno subspace. The seminal
paper by Misra and Sudarshan \cite{Misra77} showed that in general
the dynamics in the $N \to
\infty$ limit is governed by a \emph{semi}group and therefore
bears the symptoms of irreversibility. The simple finite-dimensional
example discussed in this section shows that irreversibility is not
compulsory. Notice also that in the infinite dimensional case the
meaning of $PHP$ must be defined and the self-adjointness of the
Hamiltonian $H_{\rm Z}$ cannot be taken for granted.

We focused here on \emph{one} quantum Zeno subspace ${\cal H}_P$.
However, the notion of Zeno subspace and Zeno dynamics can be
extended to a collection of projectors (an orthogonal resolution of
the identity) \cite{heraclitus}. See Sec.\
\ref{sec-nonselect}.

\section{Infinite-dimensional case: position measurement}
\label{sec-infinitedim}

We now analyze the infinite dimensional case. We will not focus on
the most general situation, but will rather study the Zeno dynamics
for the simplest spatial projection, a position measurement on a
free particle. We shall review and slightly modify the proof
given in Refs.\ \cite{FGMPS,FPSS}.

Consider a free particle of mass $m$ in $d$ dimensions
\beq
\label{eq:HU}
H=\frac{\bm p^2}{2 m}=-\frac{\Delta}{2 m}, \qquad
U(t)=\mathrm{e}^{-iHt} .
\eeq
The Hamiltonian $H$ is a positive-definite self-adjoint operator on
${\cal H}=L^2(\mathbb{R}^d)$ and $U(t)$ is unitary. Given a compact
domain $\Omega\subset\mathbb{R}^d$ with a nonempty interior and a
regular boundary, we study the evolution of the particle when it
undergoes frequent measurements defined by the projector
\beq
\label{eq:P}
P=\chi_\Omega(\bm x)=\int_\Omega \de^d \bm x \ket{\bm x}\bra{\bm x},
\qquad P
\psi(\bm x) = \chi_\Omega(\bm x) \psi(\bm x),
\eeq
where
\beq
\chi_\Omega(\bm x)=\left\{
  \begin{array}{l}
    1 \quad \mbox{for }  \bm x \in \Omega \\
    0 \quad \mbox{otherwise}
  \end{array}\right.
\label{eq:chara}
\eeq
is the characteristic function of the domain $\Omega$, and thought
of as an operator, along with its complement
$Q=1-P=1-\chi_\Omega(\bm x)$, decomposes the space
$L^2(\mathbb{R}^d)$ into two orthogonal subspaces. We study the
following process. We prepare a particle in a state with support in
$\Omega$, let it evolve under the action of its Hamiltonian, perform
frequent $P$ measurements during the time interval $[0,t]$, and
study the evolution of the system  within the Zeno subspace
${\cal H}_{P}=P{\cal H}$.

The Zeno dynamics evolution operator is given by the limit
\beq
\label{eq:UZ}
U_{\rm Z}(t) = \lim_{N\to\infty}\left[V(t/N)\right]^N,
\eeq
where the (nonunitary) evolution operator $V$ is given in Eq.\
(\ref{eq:proie}) and represents a single step
(projection-evolution-projection) Zeno process. We now show that
(\ref{eq:UZ}) yields the unitary evolution
\beq
U_{\rm Z}(t)=  \exp(-i H_{\rm Z} t) P
\label{eq:UZ1}
\eeq
generated by the Zeno Hamiltonian
\beq
\label{eq:HZ}
H_{\rm Z} =-\frac{\Delta_\Omega}{2 m},
\eeq
whose domain is a proper subspace of $L^2(\Omega)$
\beq
\label{eq:domH}
D(H_{\rm Z})=\{\psi\in L^2(\Omega)\;|\;
\Delta\psi\in L^2(\Omega), \psi(\partial \Omega)=0\},
\eeq
$\partial \Omega$ being the boundary of $\Omega$ (hard-wall or
Dirichlet boundary conditions).

One might rewrite Eq.\ (\ref{eq:HZ}) as
\beq
 H_{\rm Z}\equiv\frac{{\bm p}^2}{2m}+V_\Omega(\bm x),
\quad V_\Omega(x)=\left\{
  \begin{array}{l}
      0 \qquad \mbox{~for } \bm x \in \Omega \\
      +\infty \quad \mbox{otherwise}
  \end{array}\right. .
\label{eq:reHZ}
\eeq
In other words, the system behaves as if it were confined in
$\Omega$ by rigid walls, inducing the wave function to vanish on the
boundary of $\Omega$.

\subsection{Proof}
\label{sec-infinitedimproof}

\begin{figure}
\includegraphics[width=.6\textwidth]{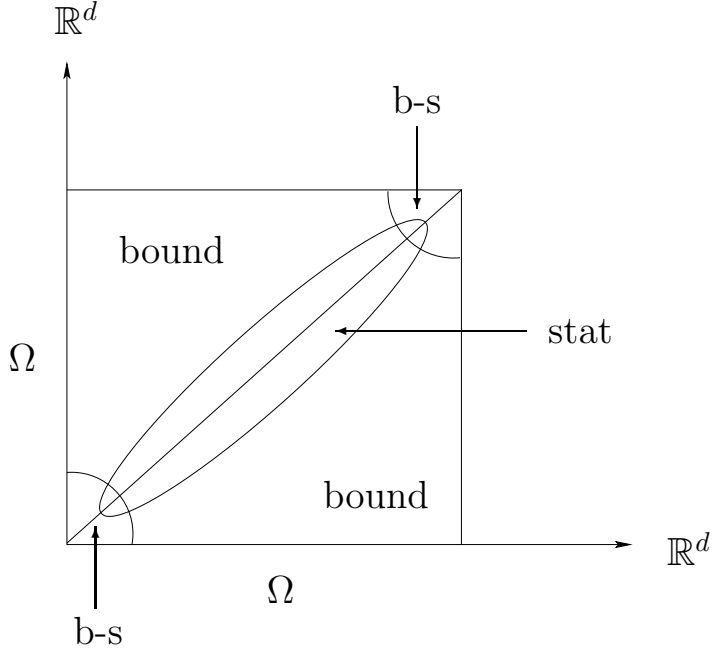}
\caption{Regularizers on the integration domain.}
\label{fig:regularizer}
\end{figure}

The matrix elements of the $d$-dimensional single-step propagator
$V$ in Eq.\ (\ref{eq:proie}) read (in the position representation)
\barr
G({\bm x},\tau; {\bm y}) & = &
\bra{\bm x} V(\tau) \ket{\bm y} \nonumber \\
& = & \chi_\Omega(\bm x) \left(\frac{m}{2\pi i
\tau}\right)^{d/2}
\exp\left[\frac{im(\bm x- \bm y)^2}{2 \tau}\right]
 \chi_\Omega(\bm y).
\label{eq:euclisinglefree}
\earr
We will work in the eigenbasis $\{\ket{{\bm n}}\}$ of $H_{\rm Z}$
belonging to the eigenvalues $E_{\bm n}$
\beq
\label{eq:eigenequ}
H_{\rm Z} \Psi_{\bm n}(\bm x) = E_{\bm n} \Psi_{\bm n}(\bm x),
\qquad \Psi_{\bm n}(\bm x)=\bra{\bm x} \bm n\rangle ,
\eeq
in the subspace $PL^2(\mathbb{R}^d)\simeq L^2(\Omega)$. This is also
the eigenbasis of $U_{\rm Z}(t)$. In this basis,
\barr
G_{\bm m ,\bm n}(\tau) &=& \bra{\bm m} V(\tau)\ket{\bm n} \nonumber \\
  &=&  \int_\Omega \de^d x\int_\Omega \de^d y
\left(\frac{m}{2\pi i \tau}\right)^{d/2}e^{i\frac{m(\bm x-\bm
y)^2}{2 \tau}} \Psi^*_{\bm m}(\bm x)\Psi_{\bm n}(\bm y)
\label{eq:Ndimprop0}
\\
&=& \int_\Omega \de^d  x \;\Psi^*_{\bm m}(\bm x)
[\mbox{bound}+\mbox{stat}] +\mbox{b-s regions},
\label{eq:Ndimprop}
\earr
where we split the integral into three parts, representing
respectively the contribution of the boundary $\bm x$ or $\bm y \in
\partial \Omega$, the stationary part $\bm x=\bm y \in \Omega - \partial
\Omega$, and the boundary points that are
\emph{also} stationary points (such points belong to the diagonal part of the
intersection of the boundaries of the two domains $\Omega$ in
(\ref{eq:Ndimprop0}), namely $\bm x = \bm y \in \partial \Omega$).
We shall separately evaluate the three contributions in the
small-$\tau$ limit, by introducing some smooth regularizing
functions and splitting the integration domain into three parts, as
shown in Fig.\
\ref{fig:regularizer}. Each regularizing function takes value one on
a compact domain and smoothly vanishes outside.

The first two terms can be evaluated by substituting $\bm \xi=\bm
y-\bm x$, to obtain
\barr
G_{\bm m , \bm n}(\tau)&=&\int_\Omega \de^d x\; \Psi_{\bm m}^*(\bm x)
\int_{\Omega-\bm x} \de^d  \xi \left(\frac{m}{2\pi i
\tau}\right)^{d/2}e^{i\frac{m \bm \xi^2}{2 \tau}}
\Psi_{\bm n}(\bm x+\bm \xi) ,
\label{eq:nnn}
\earr
where
\beq
\Omega-\bm x=\{\bm y \;|\; \bm x + \bm y \in \Omega \} .
\eeq
In order to compute the boundary term, we first observe that
\beq
e^{i \lambda \bm \xi^2} = \frac{\bm \xi \cdot \nabla e^{i \lambda
\bm \xi^2}}{2i\lambda \bm \xi^2}
\label{eq:vecfield}
\eeq
and then integrate by parts ($\lambda=m/2 \tau$)
\barr
\mbox{bound}&=& \int_\Omega \de^d \xi \left(\frac{\lambda}{\pi
i}\right)^{d/2} \Psi_{\bm n}(\bm x+\bm \xi) \frac{\bm \xi \cdot
\nabla e^{i \lambda \bm \xi^2}}{2i\lambda \bm \xi^2}
 \nonumber \\
& = & \left(\frac{\lambda}{\pi i}\right)^{d/2} \left[ \int_\Omega
\de^d
\xi\;
 \nabla\cdot \left(\frac{\Psi_{\bm n}(\bm x+\bm \xi) \bm \xi e^{i \lambda \bm \xi^2}}
{2i\lambda\bm \xi^2}\right) \right.\nonumber\\
& & \qquad\qquad
 \left. - \int_\Omega \de^d \xi\;
 \nabla\cdot \left(\frac{\Psi_{\bm n}(\bm x+\bm \xi) \bm \xi}{\bm \xi^2}\right) \frac{\bm \xi \cdot
\nabla e^{i \lambda \bm \xi^2}}{(2i\lambda)^2 \bm \xi^2} \right]
\nonumber \\
& = & \left(\frac{\lambda}{\pi i}\right)^{d/2} \left[
\oint_{\partial (\Omega -\bm x)} d^{d-1}S\;
 \frac{\Psi_{\bm n}(\bm x+\bm \xi) \bm \xi \cdot \hat{\bm u}}
{\bm \xi^2} \; \frac{e^{i \lambda \bm \xi^2}}{2i\lambda} \left(1+
O(\lambda^{-1})\right) \right]
\nonumber \\
& = & \left(\frac{m}{2\pi i \tau}\right)^{d/2}
\left[\oint_{\partial \Omega} d^{d-1}S\; \frac{\Psi_{\bm n}(\bm y) (\bm
y -\bm x) \cdot \hat{\bm u}} {(\bm y -\bm x)^2 } \; \frac{e^{i m
(\bm x -\bm y)^2/2 \tau}}{im
/ \tau }   \right. \nonumber \\
& & \qquad \qquad \qquad
\times \left(1+ \; O(\tau)\right)
\Bigg],
\label{eq:bounbb}
\earr
$\hat{\bm u}$ being the unit vector perpendicular to the boundary.
We extended the integration domain to the whole $\Omega$ and did not
explicitly write the regularizing function, that should multiply the
integrand, as its action is trivial in this case. In the second
equality, Eq.\ (\ref{eq:vecfield}) was used again in order to obtain
a higher-order volume integral with the same structure as the
initial one. Since $\Psi_{\bm n}$ in Eq.\ (\ref{eq:eigenequ}) is an
eigenfunction of $H_{\rm Z}$, whose domain is (\ref{eq:domH}), one
gets
\beq
\label{eq:bbfin}
\mbox{bound} = 0.
\eeq

The stationary contribution is obtained  by expanding the integrand
around $\bm x$
\barr
\mbox{stat}&=& \left(\frac{m}{2\pi i \tau}\right)^{d/2}
\int \de^d \xi\; e^{i \lambda \bm \xi^2} \nonumber \\
 & & \times \left(\Psi_{\bm n}(\bm  x)+
\nabla \Psi_{\bm n}(\bm x)\cdot \bm \xi+ \frac{1}{2!}
\partial_i\partial_j \Psi_{\bm n}(\bm x)\xi_i \xi_j +
O(|\bm\xi|^3)\right) .
\earr
Observe that the contributions of the linear and quadratic (with
$i\neq j$) terms in the integral vanish due to symmetry and one is
left with
\barr
\mbox{stat}&=&
\Psi_{\bm n}(\bm x)+i\frac{\tau }{2m}\Delta
\Psi_{\bm n}(\bm x)+O(\tau^{2}) \nonumber \\
& = &
\left(1-i E_{\bm
n}\tau \right)
\Psi_{\bm n}(\bm x)+O(\tau^{2}), \label{eq:statb}
\earr
where we used Eq.\ (\ref{eq:eigenequ}). Also in this case we did not
explicitly write the regularizing function in the integrand, as its
action is trivial.

Finally, we evaluate the contribution of the b-s region. Let us
first see what happens in $d=1$. We take $\Omega=[0,a]$ and compute
\beq
\mbox{b-s region}= \left(\frac{\lambda}{\pi
i}\right)^{1/2} \int_0^a \de x\int_0^a \de y e^{i\lambda (x-y)^2}
\Psi^*_{m}(x)\Psi_{n}(y)
\nu (x,y),
\label{eq:bsreg1}
\eeq
where $\nu (x,y)$ is a regularizing function that smoothly vanishes
for $|x|,|y| > \epsilon$. We expand the eigenfunctions around the
origin,
\beq
\label{eq:exppsi}
\Psi_{m}(x)= \Psi_{m}(0)+x\Psi'_m(0) + O(x^2)\sim x\Psi'_m(0) , \quad x\to 0,
\eeq
where we made use of the fact that $\Psi_{m}$ is an eigenfunction of
the Hamiltonian (\ref{eq:HZ}) and obeys Dirichlet boundary
conditions (the calculation around the other boundary point $x\simeq
y \simeq a$ is identical). Plugging into (\ref{eq:bsreg1}) and
changing integration variables $\xi = (x-y)/\sqrt{2}, \eta=
(x+y)/\sqrt{2}$, we get
\barr
\mbox{b-s region} &\sim& \left(\frac{\lambda}{\pi
i}\right)^{1/2} \int_0^\epsilon \de x\int_0^x \de y \; x y \;
e^{i\lambda (x-y)^2} \Psi^{*'}_{m}(0)\Psi'_{n}(0) \nonumber \\
&=&
\left(\frac{\lambda}{\pi
i}\right)^{1/2} \Psi^{*'}_{m}(0)\Psi'_{n}(0)
\int_0^{\epsilon/\sqrt{2}} \de \eta
\int_0^\eta \de \xi
 e^{i2\lambda \xi^2}
(\eta^2 - \xi^2)
\nonumber \\
& = &
\left(\frac{1}{\pi i}\right)^{1/2} \Psi^{*'}_{m}(0)\Psi'_{n}(0)
\left(\frac{\sqrt{\pi i}}{24} \epsilon^3  -
\frac{i^{3/2}}{16\lambda}\epsilon -\frac{1}{12\lambda^{3/2}}
+ o(\lambda^{-3/2})\right)
\nonumber \\
&=& O(\tau^{3/2}) + \epsilon O(\tau) + \frac{\epsilon^3}{24},
\label{eq:bsa}
\earr
and by sending $\epsilon \to 0$ we obtain
\beq
\mbox{b-s region} = O(\tau^{3/2}).
\label{eq:bsa}
\eeq
Note that the residual $\epsilon O(\tau)$ contribution is due to the
stationary points in $[0,\epsilon]^2$  and belongs to the bulk of
Eq.\ (\ref{eq:statb}).

In $d$ dimensions the proof is similar. By writing
\barr
\mbox{b-s region} &=& \left(\frac{\lambda}{\pi
i}\right)^{d/2} \int_{\Omega} \de^d x\int_{\Omega}
\de^d y e^{i\lambda (\bm x-\bm y)^2} \Psi^*_{\bm m}(\bm x)\Psi_{\bm
n}(\bm y) \nu (\bm x , \bm y ), \nonumber \\
 &\simeq& \left(\frac{\lambda}{\pi
i}\right)^{d/2} \int_{\omega_\epsilon} \de^d x\int_{\omega_\epsilon}
\de^d y e^{i\lambda (\bm x-\bm y)^2} \Psi^*_{\bm m}(\bm x)\Psi_{\bm
n}(\bm y) ,
\label{eq:bsregd}
\earr
where $\nu (\bm x, \bm y)$ is the regularizing function and
\beq
\omega_\epsilon
=\{\bm x \in \Omega | d(\bm{x},\partial\Omega)<\epsilon\} ,
\qquad d(\bm{x},\partial\Omega) = \inf_{\bm{y}\in\partial\Omega}|\bm{x}-\bm
{y}| ,
 \label{eq:bsregdfin}
\eeq
one obtains
\beq
\mbox{b-s region}= o(\tau) .
\label{eq:bsfin}
\eeq
By plugging (\ref{eq:bbfin}), (\ref{eq:statb}) and (\ref{eq:bsfin})
into (\ref{eq:nnn}) we obtain the matrix elements of the single-step
operator
\beq
\label{eq:diag}
G_{\bm m ,\bm n}(\tau) =\delta_{\bm m, \bm n}\left(1-i E_{\bm
n}\tau \right)+ R_{\bm m ,\bm n}(\tau),
\eeq
where for $\tau\to0$
\beq
\label{eq:diagbis}
R_{\bm m,\bm n}(\tau)=o(\tau),
\eeq
and under the assumption of uniform convergence of the infinite sums
stemming from the insertion of $N-1$ resolutions of the identity in
(\ref{eq:UZ}), one obtains ($\tau=t/N$):
\barr
\label{eq:new}
G^{\rm Z}_{\bm m ,\bm n}(t)  &=&
\bra{\bm m}U_Z(t)\ket{\bm n} \nonumber \\
&=&\lim_{N \to \infty} \sum_{\bm n_1,\ldots, \bm n_{N-1}} G_{\bm
m, \bm n_1}(t/N)G_{\bm n_1,\bm n_2}(t/N) \cdots G_{\bm n_{N-1},\bm
n}(t/N)
\nonumber\\
&=& \delta_{\bm m, \bm n} \exp\left(-i E_{\bm n}
t\right).
\earr
This is precisely the propagator of a particle in a box with
Dirichlet boundary conditions. This in turn proves that $H_{\rm Z}$
is given in (\ref{eq:HZ}) and has eigenbasis $\{\ket{\bm n}\}$. Note
also that the $o(t)$ contribution (\ref{eq:diagbis}) drops out of
(\ref{eq:new}) in the $N\to\infty$ limit since it appears as $N
\times o(1/N)$.

\subsection{A few comments on the proof}
\label{sec-potentialzenoa}

It is worth emphasizing that the basis given in Eq.\
(\ref{eq:eigenequ}) is only one of many (infinite in fact)
possibilities for a basis for the domain $\Omega$. Any one of these
would be valid, but not all would be equally convenient. Thus with a
basis whose functions did not vanish at the boundary $\partial
\Omega$, the dominant contribution of order $\lambda^{-d/2}$ in
the function bound$(\lambda)$ in (\ref{eq:bounbb}) would have given
a nondiagonal term both in (\ref{eq:bounbb}) and (\ref{eq:new}). The
matrix representation of $G$ (in this basis) would in that case
still need to be diagonalized, leading back to the matrix we have
found using a more convenient basis. Our point is that one can
always choose to use the basis $\{\ket{\bm n}\}$ of
(\ref{eq:eigenequ}). For that choice the calculation is easiest and
the resulting interpretation transparent.

Note also that in the preceding proof the detailed features of the
convergence of the limits are not worked out. We implicitly assumed
the uniform convergence of the infinite sums in Eq.\ (\ref{eq:new}).
Much additional care is required at a rigorous mathematical level,
where one must prove that the limits can be interchanged. We shall
reconsider this problem in much greater details in the following
sections.

\subsection{Particle in a potential}
\label{sec-potentialzeno}

The introduction of a potential is not difficult to deal with if
mathematical subtelties are not spelled out. Let us therefore
proceed formally and extend the proof that spatial projections yield
ordinary constraints (Dirichlet) when the particle moves in a
sufficiently regular potential. The situation clearly becomes more
complicated when the potential is singular and/or the projected
spatial region (or its boundary) lacks the required regularity.

Let
\andy{HamV}
\beq
H=\frac{{\bm p}^2}{2m}+V (\bm x),\qquad U(t)=\exp(-i t H ),
\label{eq:HamV}
\eeq
where $V$ is a regular potential. (It may be unbounded from below,
for example $V(\bm x)=F x_j $ for some $j$, although within the
projected region $\Omega$ the total Hamiltonian $H$ should be lower
bounded.) The measurement is again application of the projector
(\ref{eq:chara}) and we simply replace the short-time propagator
(\ref{eq:euclisinglefree}) with
\andy{euclisingleV}
\barr
G({\bm x},\tau; {\bm y})  &=& \chi_\Omega(x) \left(\frac{m}{2\pi
i \tau}\right)^{d/2}\exp\left[\frac{im({\bm x}-{\bm
y})^2}{2
\tau}\right] \nonumber \\
& & \qquad \qquad \times \exp\left[-\frac{i\tau(V({\bm x})+V({\bm
y}))}{2}\right]
\chi_\Omega(y).
\label{eq:euclisingleV}
\earr
We make use again of the eigenbasis of the Hamiltonian with
Dirichlet boundary conditions on $\Omega$
\andy{baseV}
\beq
H_{\rm Z}|\Psi_{\bm n}\rangle = \left(\frac{{\bm p}^2}{2m}+V({\bm
x}) \right) |\Psi_{\bm n}\rangle = E_n |\Psi_{\bm n}\rangle\, ,
 \quad \Psi_n({\bm x})|_{{\bm x}\in \partial \Omega}=0
\label{eq:baseV}
\eeq
and notice that the eigenfunction can be expanded as in
(\ref{eq:exppsi}) by virtue of the regularity of the potential. A
calculation identical to the previous one yields
\beq
\label{eq:diagaa}
G_{\bm m ,\bm n}(\tau) =\delta_{\bm m, \bm n}\left(1-i E_{\bm
n}\tau\right)+ R_{\bm m ,\bm n}(\tau),
\eeq
where again $R_{\bm m,\bm n}(\tau)=o(\tau)$, so that
\barr
\label{eq:newww}
G^{\rm Z}_{\bm m ,\bm n}(t) =
\bra{\bm m}U_{\rm Z}(t)\ket{\bm n}
&=& \delta_{\bm m, \bm n} \exp\left(-i  E_{\bm n}
t\right).
\earr
In conclusion, the evolution in the Zeno subspace is governed by the
Hamiltonian
\andy{HZV}
\beq
H_{\rm Z}=\frac{{\bm p}^2}{2m}+V_\Omega({\bm x}),
\quad V_\Omega( {\bm x})=\left\{
  \begin{array}{l}
    V({\bm x}) \quad \mbox{for } {\bm x} \in \Omega \\
    +\infty \quad \mbox{otherwise}
  \end{array}\right.
\label{eq:HZV}
\eeq
We notice here something interesting. We need only require that the
Hamiltonian be lower bounded in the Zeno subspace. Although for
unbounded potentials (like $V=F x_j$) $H$ may not be lower bounded,
$V_\Omega({\bm x})$ can be lower bounded in $\Omega$, yielding
unitary evolution operators.

\subsection{The physics behind the ``hard wall"}
\label{sec-comdbc}
If we ponder over the proofs of this section, we understand how the
Zeno mechanism prevents leakage out of the Zeno subspace. Frequent
projections force the wave function to vanish on the boundary of the
spatial region associated with the projection. In turn, this implies
a vanishing current through the boundary. This is equivalent
to a ``hard wall". The derivation of the Dirichlet boundary
conditions has implications for this notion, as used for example in
elementary quantum mechanics. Everyone would agree  that
this notion is an idealization. However, in many cases where this
idealization is useful the ``wall" is dynamic rather than static,
the result of some fluctuating atomic presence. We have here a
sufficient condition for the validity of this notion in a dynamic
situation. Moreover, there is a quantitative framework (arising from
our asymptotic analysis and finite-time-interval QZE effects) for
gauging the effects of less than perfect hard walls. As we will see
in Sec.\ \ref{sec:tuttoins}, this has also spinoffs for the notion
of constraint in quantum mechanics.

\subsection{Algebra of observables in the Zeno subspace and Zeno dynamics in Heisenberg picture}
 \label{sec-algobs}

We now look at the Zeno dynamics in the Heisenberg picture. The
following discussion is an exploratory investigation. A natural
question concerns the destiny of the algebra of observables after
the projection \cite{FMPSS}. This is not a simple problem. One can
assume that to a given observable $\mathcal{O}$ before the Zeno
projection procedure there corresponds the observable
$P\mathcal{O}P$ in the projected space:
\beq
\mathcal{O} \Rightarrow P\mathcal{O}P.
\label{eq:OPOP}
\eeq
For example, if one starts in $\mathbb{R}$ and projects over a
finite interval $I$ of $\mathbb{R}$, $P=\chi_I(x)$, the momentum and
position operators become
\barr
p \Rightarrow PpP = \left\{ \matrix{ i\partial_x & \mbox{for} \; x
\in I \cr
              0  & \mbox{otherwise}}  \right. \ ,\\
x \Rightarrow PxP = \left\{ \matrix{x & \mbox{for} \; x \in I \cr
              0 & \mbox{otherwise}}\right. \ .
\earr
Observe that the correspondence (\ref{eq:OPOP}) is \emph{not} an
algebra homomorphism. However, if we redefine a new associative
product in the algebra of operators, by setting
\beq
A*B \equiv APB ,
\label{eq:star}
\eeq
with this new product the previous correspondence (\ref{eq:OPOP})
becomes an algebra homomorphism \cite{MMSZ,carinena}. Notice also
that the new (projected) algebra acquires a unity operator $P$. In
general the evolution will not be an automorphism of the new
product. However, it will respect the product to order $O(t/N)$ and
induce, in the limit, a Zeno dynamics on the projected algebra,
i.e.\ on the image of the projection. The evolution will be
trivially an automorphism when it commutes with $P$ and is therefore
compatible with the new product without any approximation.

In general one has to modify the associative product in such a way
that the ``deviation"  of $U(t/N)$ from being an automorphism is of
order $o(t/N)$, so that in the limit $U_{\rm Z}(t)$ will be an
automorphism of the new associative product adapted to the
constraint. In other words, the sequence of evolution operators
\begin{equation}\label{eq:evseq}
V_N(t)=V(t/N)^N=(P U(t/N) P)^N ,
\end{equation}
yielding the Zeno limit (\ref{eq:UZ}), is mirrored at the level of
the algebra by the following sequence of deformed associative
products
\begin{equation}\label{eq:prodseq}
A *_N B \equiv A P_N B ,
\end{equation}
where $P_N$ is a  positive operator with $0\leq P_N\leq 1$ and
$P_N P = P P_N= P$. For any $N$, $P_N$ forms together with
$Q_N=1-P_N$ a positive operator valued measure, yielding a
resolution of the identity, i.e.\ $P_N+Q_N=1$, which approximates
the orthogonal resolution $P+Q=1$, in the sense that
\begin{equation}\label{eq:PNord}
P_N \psi =P \psi +O(1/N) , \qquad \forall \psi\in \cH
.
\end{equation}
For any $N$ the evolution $V_N(t)$ is an automorphism of the product
$*_N$ and in the limit $N\to\infty$ we get the desired result
(\ref{eq:star}).

Observe that, for unbounded operators, (\ref{eq:OPOP}) does not
necessarily yield self-adjoint operators: for example, after the
Zeno procedure, the momentum $p$ would act on functions that vanish
on the boundary of $I$ and would have deficiencies $\langle 1,1
\rangle$, see \cite{FGMPS}. On the other hand the Zeno Hamiltonian
(\ref{eq:HZ}) is self-adjoint. However, it would be arbitrary to
require a similar property for every observable in the algebra. In
general, we speculate that the lack of self-adjointness of the
operators representing the ``observables" of the system in the
projected subspace might be related to the incompleteness of the
corresponding classical field
\cite{nelson,Klauder,FGMPS}.

\subsection{Projections onto lower dimensional regions: constraints}
\label{sec:tuttoins}

In all the situations considered so far, the projected domain always
has the same dimensionality of the original space ($\mathbb{R}^n$).
[Remember that, after Eq.\ (\ref{eq:HU}), we required the projected
domain $\Omega$ to have a nonempty interior.] However, it is
interesting to ask what would happen if one would project onto a
domain $\Omega'$ of lower dimensionality \cite{FMPSS}. This is
clearly a more delicate problem, as one necessarily has to face the
presence of divergences. It goes without saying that these
divergences must be ascribed to the lower dimensionality of the
projected domain and \emph{not} directly to the convergence features
of the Zeno propagator \cite{Froese}. Our problem is to understand
how these divergences can be cured. One way to tackle this problem
is to start from a projection onto a domain $\Omega
\subset \mathbb{R}^n$ and {\it then} take the limit $\Omega \to
\Omega'\subset \mathbb{R}^{n-1}$, with a Hilbert space (Zeno subspace)
$L^2(\Omega')$ \cite{FMPSS}.

These problems are still very open (even at the level of formal
derivations) and lead us to interesting links with constrained
dynamics in quantum mechanics and quantum field theory. Curiously,
the Zeno phenomenon and the Zeno dynamics might suggest strategies
in order to impose constraints onto quantum evolutions.

\section{Bounded Hamiltonians}
\label{sec-bounded}

So far, for the sake of simplicity and illustration, our analysis
lacked mathematical rigor. The present and the following seven
sections will have a different character. We shall focus on the
conditions that must be required in order that the analysis be
mathematically sound.

Consider a bounded Hamiltonian $H$, with  $H=H^\dagger$ and  $\|H\| < \infty$. The one
parameter unitary group
\begin{equation}
U(t)=\exp(-i H t) = \sum_{n\geq 0} \frac{(-i t)^n}{n!} H^n
\label{eq:boundedexp}
\end{equation}
is uniformly continuous, with norm derivative $U'(0)=-i H$, that is
$\lim_{s\to t}\|U(s)-U(t)\|=0$ and $\lim_{s\to 0}\|(U(s)-U(0))/s - i H\|=0$ \cite{Kato}. In
this case it is very easy to prove the existence and explicitly
derive the expression of the (uniform) limit of the Zeno product
formula
\begin{equation}
U_{\mathrm{Z}}(t) =\lim_{N\to\infty} V_N(t), \qquad V_N(t)=\left[P
U\left(\frac{t}{N}\right) P \right]^N.
\label{eq:ZPF}
\end{equation}
Indeed, by the existence of the norm derivative, or directly by
(\ref{eq:boundedexp}),
\begin{equation}
U(t)=1-i H t + o(t),
\end{equation}
where $o(t)$ is an operator valued function defined in a
neighborhood of $0$, such that $\|o(t)\|/t \to 0$ as $t\to 0$.
Therefore,
\begin{eqnarray}
V_N(t) &=& \left[P \left(1-i H \frac{t}{N}
+o\left(\frac{t}{N}\right) \right)P \right]^N
\nonumber\\
&=& P \left[1-i P H P \frac{t}{N} + o\left(\frac{t}{N}\right)
\right]^N .
\end{eqnarray}
By using the following straightforward equality
\begin{equation}
\left[1+ \frac{A}{N} + o\left(\frac{1}{N}\right)\right]^N =
\sum_{k=0}^N \frac{A^k}{k!} + o(1) \stackrel{N\to\infty}{\longrightarrow} \exp A,
\end{equation}
valid for any bounded $A$, one obtains the desired result
\begin{equation}
U_{\mathrm{Z}}(t) = P \exp\left(-i H_{\mathrm{Z}} t\right),
\qquad
H_{\mathrm{Z}} = PHP ,
\end{equation}
uniformly in any compact $t$ interval. The Zeno dynamics is thus
rigorously proved for a bounded Hamiltonian. Although not explicitly
stated, the result of Sec.\ \ref{sec-finitedim} is a particular case
of the above: for finite dimensional systems, $H$ is bounded and the
result rigorous.

\subsection{Examples}
\label{sec-boundedex}

A first (somewhat trivial) example is $H=$ finite-rank operator (a
matrix; remember that the Hilbert space is in general
infinite-dimensional). Then $H_{\mathrm{Z}} = PHP$ is block
diagonal.

As a second example, consider a particle on the real line and take
the Hamiltonian
\begin{equation}
H =  p^2 e^{- \Lambda ^2 p^2} ,
\end{equation}
$p$ being the momentum operator and $\Lambda> 0$ a cutoff, and the projection
\begin{equation}
P=\chi_{\mathbb{R}_+}(x),
\end{equation}
$\chi$ being the characteristic function. A Zeno effect takes place
and the Zeno dynamics on the positive half-line is governed by the
operators
\begin{equation}
U_{\mathrm{Z}}(t) = P \exp\left(-i H_{\mathrm{Z}} t\right),
\qquad
H_{\mathrm{Z}} = P p^2 e^{- \Lambda ^2 p^2} P .
\end{equation}

\section{Unbounded Hamiltonians}
\label{sec-unbound}
Let us now consider the case of an unbounded Hamiltonian $H$. In
such a situation there are two serious problems that must be faced:
the existence of the limit of the Zeno product formula
(\ref{eq:ZPF}) and the form of the limiting dynamics. In particular,
one can ask under which conditions the (limiting) Zeno dynamics
exists and under which additional conditions it is a unitary group
in the Zeno subspace. This problem will occupy us for the next few
sections.

\begin{figure}
\includegraphics[width=\textwidth]{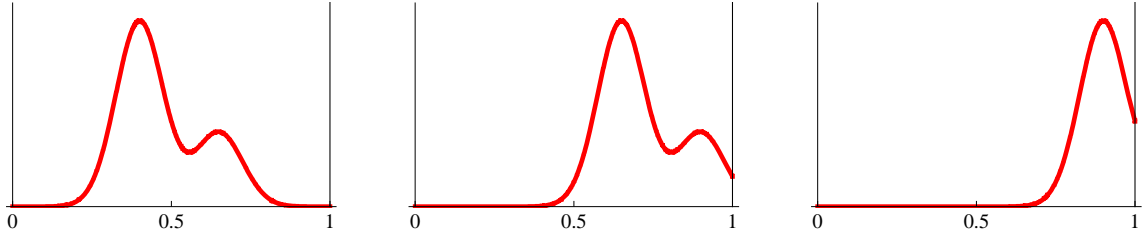}
\caption{Zeno dynamics of the momentum operator.}
\label{fig:compact}
\end{figure}
Let us start from an example and show that in order to obtain a
unitary group, one should restrict one's attention to semibounded
Hamiltonians. Consider the right translation on the line. The
Hilbert space is $\mathcal{H}=L^2 (\mathbb{R})$ and the Hamiltonian
is taken to be the momentum operator $H=p=-i\partial_x$ with domain
$D(H)=H^1(\mathbb{R})$, where $H^1(\mathbb{R})=\{\varphi\in
L^2(\mathbb{R}) |\partial_x\varphi\in L^2(\mathbb{R}) \}$ is the
Sobolev space. Note that $p$ is self-adjoint and unbounded both below
and above, since its  spectrum is the whole line $\mathbb{R}$. Let
us choose the projection on the unit segment $P=\chi_{[0,1]}(x)$, so
that $\mathcal{H}_P = P \mathcal{H}\simeq L^2(0,1)$. We get
\begin{equation}
\mathrm{e}^{-i t p}\chi_{[0,1]}(x) \mathrm{e}^{i t p}=
\chi_{[0,1]}(x-t)= \chi_{[t,1+t]} (x),
\end{equation}
hence, for any $t,s\in \mathbb{R}$ with $t s \geq 0$,
\begin{equation}
\chi_{[0,1]} \mathrm{e}^{-i t p}\chi_{[0,1]}\mathrm{e}^{-i s p}\chi_{[0,1]}=
\chi_{[0,1]} \chi_{[t,1+t]} \mathrm{e}^{-i (t+s) p}\chi_{[0,1]}.
\end{equation}
On the other hand,
\begin{equation}
\mathrm{e}^{-i (t+s) p}\chi_{[0,1]}= \chi_{[t+s,1+t+s]} \mathrm{e}^{-i (t+s) p}\chi_{[0,1]}
\end{equation}
and thus
\begin{eqnarray}
\chi_{[0,1]} \mathrm{e}^{-i t p}\chi_{[0,1]}\mathrm{e}^{-i s p}\chi_{[0,1]} &=& \chi_{[0,1]} \chi_{[t,1+t]} \chi_{[t+s,1+t+s]} \mathrm{e}^{-i (t+s) p}\chi_{[0,1]} \nonumber\\
&=& \chi_{[0,1]} \mathrm{e}^{-i (t+s) p}\chi_{[0,1]} ,
\end{eqnarray}
that is
\begin{equation}
(P  \mathrm{e}^{-i t H} P)(P \mathrm{e}^{-i s H} P) = P  \mathrm{e}^{-i (t+s) H} P,
\end{equation}
when $t s\geq 0$.

Therefore, since the Zeno product formula does not depends on $N$
\begin{equation}
V_N(t)= P \mathrm{e}^{-i t H} P, \qquad N\geq1,
\end{equation}
its limit exists and reads
\begin{equation}
V_{\mathrm{Z}}(t)= P \mathrm{e}^{-i t H} P.
\end{equation}
The Zeno dynamics is represented in Fig.\ \ref{fig:compact} and is
clearly not unitary in $\mathcal{H}_P$. Rather, it is a contractive
semigroup describing probability leakage out of the Zeno subspace.
In a sense, in this situation there is no quantum Zeno effect.

In the following we will therefore restrict our attention to
semibounded operators, and, for definiteness, to positive
Hamiltonians. This entails no loss of generality, because any
semibounded operator $H$  can be written as $H=\pm H_0 + c$, with
$c\in\mathbb{R}$ and $H_0\geq 0$.

\section{One dimensional projection}
\label{sec-onedim}

Let us start by considering a one dimensional projection $P$ with
range $\cH_P=\mathrm{span}\{ \psi \}$ (where for simplicity we write
$\psi=\psi_0=$ initial state) and a one parameter group of unitaries
$\{U(t)
\}_t$ with a (generally unbounded) positive generator $H\geq0$,
\begin{equation}
P= \ket{\psi} \bra{\psi}, \qquad U(t)= \exp(-i H t)  .
\end{equation}
The Zeno product formula reads
\begin{equation}
\label{eq:Zenoprod}
V_N(t)=\left[P U\left(\frac{t}{N}\right) P \right]^N= P
\left[\bra{\psi}U\left(\frac{t}{N}\right)\ket{\psi}\right]^N
\end{equation}
Therefore, one has to study the limit
\begin{equation}
F(t)=\lim_{N\to+\infty}  \left[\As\left(\frac{t}{N}\right)\right]^N,
\qquad
\As(t)=\bra{\psi}U(t)\ket{\psi} .
\end{equation}
By noting that $\As(0)=1$ one gets
\begin{equation}
F(t)= \exp\left(t \As'(0)\right)
\end{equation}
where
\begin{equation}
\As'(0)=\lim_{s\to 0} \frac{\As(s)-1}{s}=\lim_{s\to 0}
\bra{\psi}\frac{U(s)-1}{s}\ket{\psi}.
\end{equation}
If   $\psi \in D(H^{1/2})$, then the above limit exists  and reads
\begin{equation}
\label{eq:A'(0)}
\As'(0)=-i \left\| H^{1/2} \psi \right\|^2=-i \bra{H^{1/2} \psi}\left.
H^{1/2} \psi\right\rangle.
\end{equation}
The proof is easily given in terms of the spectral representation
$E(\lambda)$ of $H$: from
\begin{equation}
H=\int_0^{+\infty} \lambda \de E(\lambda), \qquad
H^{1/2}=\int_0^{+\infty} \lambda^{1/2} \de E(\lambda),
\end{equation}
one gets
\begin{equation}
 \bra{\psi}\frac{U(s)-1}{s}\ket{\psi}=\int_0^{+\infty}
 \frac{\mathrm{e}^{-i\lambda s}-1}{s} \de \|E(\lambda)\psi\|^2 ,
\end{equation}
where $\de \|E(\lambda)\psi\|^2 = \de \langle \psi, E(\lambda)\psi
\rangle$. If $\psi \in D(H^{1/2})$, i.e.
\begin{equation}
\left\| H^{1/2} \psi \right\|^2=\int_0^{+\infty} \lambda \de \|E(\lambda)\psi\|^2 < \infty,
\end{equation}
then by noting that
\begin{equation}
\left|\frac{\mathrm{e}^{-i\lambda s}-1}{s}\right|=\left|\lambda\,
\mathrm{sinc}\left(\frac{\lambda s}{2}\right)\right| \leq |\lambda| ,
\end{equation}
by dominated convergence one gets  (\ref{eq:A'(0)}) .
Therefore
\begin{equation}
U_{\mathrm{Z}}(t) =\mathrm{s-}\!\!\!\lim_{N\to\infty} V_N(t)=
P\exp\left(-i \left\| H^{1/2} \psi \right\|^2 t \right)
\end{equation}
$\forall t\in \mathbb{R}$ and uniformly in any compact interval.
Here, $A=\mathrm{s-}\lim_{N} A_N$ denotes the strong operator limit,
that is $\lim_N A_N \varphi = A \varphi$, $\forall\varphi\in\mathcal{H}$.
In fact, in this case the limit holds in norm, for
\begin{equation}
\left\| V_N(t)  -U_{\mathrm{Z}}(t)\right\| = \left| \As\left(\frac{t}{N}\right)^N- F(t)\right|\| P\| \to 0,
\qquad N\to+\infty .
\end{equation}
Therefore, the (trivial) evolution in the one-dimensional subspace
$\cH_P$ is engendered by the phase $ \left\| H^{1/2} \psi
\right\|^2$, that is by the Zeno Hamiltonian
\begin{eqnarray}
H_{\mathrm{Z}}&=& \left\| H^{1/2} \psi \right\|^2 P =
\ket{\psi}\bra{H^{1/2}  \psi} \left. H^{1/2}
\psi\right\rangle\bra{\psi}
\nonumber\\
&=&(H^{1/2} P)^\dagger (H^{1/2} P).
\end{eqnarray}
Incidentally, $H^{1/2} P$ -- and thus $(H^{1/2} P)^\dagger$ -- is a
bounded operator, with $D(H^{1/2} P)=\cH$, for
\begin{equation}
\left\|  H^{1/2} P \right\|=\left\|  H^{1/2} \psi \right\|< \infty .
\end{equation}
In conclusion, for a one dimensional projection $P$, the limit of
the Zeno product formula (\ref{eq:Zenoprod}) exists if
$D(H^{1/2}P)=\cH$ and is given by
\begin{eqnarray}
U_{\mathrm{Z}}(t) &=&\mathrm{s-}\!\!\!\lim_{N\to\infty} V_N(t)=
P\exp\left(-i H_{\mathrm{Z}} t\right)
\nonumber\\
&=& P\exp\left(-i   (H^{1/2} P)^\dagger (H^{1/2} P) t\right) .
\end{eqnarray}
Moreover, the limit holds in norm, uniformly in $t$ in any compact
subset of $\mathbb{R}$.

\subsection{Example}
\label{sec-unboundedex}

Consider a free particle on the real line, $\cH=L^2(\mathbb{R})$,  and take the Hamiltonian
\begin{equation}
H = p^2 ,
\end{equation}
$p$ being the momentum operator. Let the measurement be associated
to the projection operator $P=\ket{\psi}\bra{\psi}$, that projects
the system onto the state
\begin{equation}
\hat \psi(p) = \bra{p}\psi\rangle = \frac{N}{a+|p|^{5/2}}\; ,
\end{equation}
where $a$ is a positive constant and $N$ a normalization factor.
This state does not belong to the domain of the Hamiltonian,
\begin{equation}
\|H \psi \|^2=
\int p^4 |\hat \psi(p)|^2 \de p= \infty.
\end{equation}
However, it belongs to the domain of $H^{1/2}$:
\begin{equation}
\|H^{1/2} \psi \|^2=\int p^2 |\hat \psi(p)|^2 \de p = E_\psi < \infty.
\end{equation}
A Zeno effect takes place and the Zeno Hamiltonian reduces to a
phase:
\begin{equation}
H_{\mathrm{Z}} = E_\psi P .
\end{equation}

\section{Finite dimensional projection}
\label{sec-findim}
The generalization to finite dimensional projections is
straightforward. First notice that the sufficient condition $\psi\in
D(H^{1/2})$ translates into $\cH_P\subset D(H^{1/2})$, i.e.\
$D(H^{1/2}P)=\cH$. In fact, $H^{1/2}P$ is not only bounded, but also
a finite rank operator. Therefore all the results of the previous
subsection immediately translate into analogous results.

Consider a finite dimensional projection $P$ and a positive
Hamiltonian $H$. If $D(H^{1/2}P)=\cH$ the limit of the Zeno product
formula (\ref{eq:Zenoprod}) exists and is given by
\begin{eqnarray}
U_{\mathrm{Z}}(t) =\mathrm{s-}\!\!\!\lim_{N\to\infty} V_N(t)=
P\exp\left(-i H_{\mathrm{Z}} t\right),
\end{eqnarray}
where
\begin{equation}
\label{eq:Hzzz}
 H_{\mathrm{Z}}= (H^{1/2} P)^\dagger (H^{1/2} P).
\end{equation}
Moreover, the limit holds in norm, uniformly in any bounded interval
of $t$.

Note that all the experiments performed so far make use of finite
dimensional projections (onto a finite numbers of quantum levels)
and belong to this class. Moreover, observe that in this simple case
one is able to give a precise mathematical meaning to the physical
intuition that the limiting Zeno Hamiltonian must be $PHP$. As a
matter of fact,
 $H_{\mathrm{Z}}$ in Eq.\ (\ref{eq:Hzzz}) is
nothing but the corresponding rigorous expression.  Note also that,
in general, $PHP\subset (H^{1/2} P)^\dagger (H^{1/2} P)$ as a proper
restriction, but if $\cH_P\subset D(H) \subset D(H^{1/2})$, i.e.\
$D(HP)=\cH$, the Zeno Hamiltonian simplifies into
\begin{equation}
H_{\mathrm{Z}} =P H P .
\end{equation}
Obviously, the last condition is always satisfied for bounded $H$
and the results of Sec.\
\ref{sec-bounded} are reobtained.

\section{Product formulae}

We now study more general product formulae, clarifying what is the
state of the art and what can be said when the Hamiltonian is
unbounded \emph{and} the projection operator
infinite dimensional. The general mathematical problem is still open
and of great interest.
We start this
section with a formula due to Trotter, in which no projection
operators appear, and then partially extend these results to the
Zeno dynamics.

\subsection{Trotter}
\label{sec-trotter}
\andy{trotter}

Let $A$ and $B$ be self-adjoint operators with domains $D(A)$ and
$D(B)$ and let $A+B$ be essentially self-adjoint on $D(A+B)=D(A)\cap
D(B)$. Then  \cite{Trotter1,Trotter2}
\beq
\mathrm{s-}\!\!\!\lim_{N\to\infty}  \left(\mathrm{e}^{i A t/N}  \mathrm{e}^{i B t/N} \right)^N = \mathrm{e}^{i(A+B)t}
\label{eq:trotterf}
\eeq
for all $t\in\mathbb{R}$, uniformly on compact sets. Moreover, if
$A$ and $B$ are lower bounded, then
\beq
\mathrm{s-}\!\!\!\lim_{N\to\infty}  \left(\mathrm{e}^{-A t/N}  \mathrm{e}^{-B t/N} \right)^N = \mathrm{e}^{-(A+B)t}
\label{eq:trotterf1}
\eeq
for all $t\geq 0$, uniformly on compact sets. This is the
celebrated Trotter product formula.

Recall that a symmetric operator $T$ (that is, a densely defined
operator with $T\subset T^\dagger$) is said to be essentially
self-adjoint if its closure $T^{\dagger\dagger}$ is self-adjoint.

\subsection{Kato}
Let $A$ and $B$  be positive self-adjoint operators in $P_A\cH$ and $P_B\cH$, where
$P_A$, $P_B$ are the projections on the closures of $\overline{D(A)}$ and $\overline{D(B)}$, respectively.
Let $D=D(A^{1/2})\cap D(B^{1/2})$ and let $P_C$ be the
projection on $\overline{D}$.
 Then one gets \cite{Kato1}
\begin{equation}
\mathrm{s-}\!\lim_{N\to\infty}\left[ \exp(-A t/N) P_A \exp(-B t/N) P_B\right]^N=  \exp(-t C) P_C,
\label{eq:Kato}
\end{equation}
for all $t\geq 0$, where $C$ is the form sum of $A$ and $B$, i.e.\
the self-adjoint operator in $P_C\cH$ associated with the closed
densely defined quadratic form $\mathfrak{q}$ on $D$:
$\mathfrak{q}(\varphi)=\|A^{1/2}\varphi \|^2 +\|B^{1/2}\varphi
\|^2$.
This formula is due to Kato, who gave important contributions
in this field and motivated many studies by several authors.

\subsection{Corollary: Self-adjoint Zeno product formula}
Note that if $A=0$, $P_A=P$, $B=H$ and $P_B=1$ one gets $D=\cH_P\cap
D(H^{1/2})=\cH_P\cap D(H^{1/2}P)$. Now, if $D(H^{1/2}P)$ is dense in
$\cH$, i.e.\ $\overline{D}=\cH_P$, one also gets $P_C=P$. Therefore,
Kato's formula (\ref{eq:Kato}) translates into
\begin{equation}
\mathrm{s-}\!\lim_{N\to\infty}\left[  P \exp(-H t/N) \right]^N= \exp(-t H_{\mathrm{Z}}) P,
\qquad t\geq 0,
\label{eq:ZenoKato}
\end{equation}
where $H_{\mathrm{Z}}$ is the Zeno Hamiltonian, associated with
$\mathfrak{h}_{\mathrm{Z}}(\varphi)=\|H^{1/2} P\varphi \|^2$, i.e.\
$H_{\mathrm{Z}}=(H^{1/2} P)^\dagger (H^{1/2} P)$. Note also that
\begin{equation}
\left[  P \exp(-H t/N) P \right]^N= \left[  P \exp(-H t/N) \right]^N P ,
\end{equation}
and the symmetric \emph{self-adjoint} Zeno product formula follows:
if $H\geq 0$ and $D(H^{1/2}P)$ is dense in $\cH$, then
\begin{equation}
\mathrm{s-}\!\lim_{N\to\infty}\left[  P \exp(-H t/N)  P\right]^N= P \exp(-t H_{\mathrm{Z}}) ,
\qquad t\geq 0,
\label{eq:ZenoKatosymm}
\end{equation}
with
\begin{equation}
\label{eq:HZinter}
H_{\mathrm{Z}}=(H^{1/2} P)^\dagger (H^{1/2} P).
\end{equation}
This is the correct self-adjoint extension of the ``physical"
Hamiltonian $PHP$.

\section{The theorem of Misra and Sudarshan}

\label{sec-MStheorem}
\andy{MStheorem}

In order to investigate the structure of the Zeno limit, Misra and
Sudarshan \cite{Misra77} completely bypass the problem of its
existence. Instead, they assume that it exists.  Consider
\begin{equation}
V_N(t)=\left[P U\left(\frac{t}{N}\right) P \right]^N,
\end{equation}
with $H>0$ unbounded and $P$ infinite dimensional projection. Assume
that
\begin{equation}
U_{\mathrm{Z}}(t) =\mathrm{s-}\!\!\!\lim_{N\to\infty}  V_N(t)
\label{eq:ZenolimVN}
\end{equation}
exists for all $t\in\mathbb{R}$ and that it is strongly continuous
at $t=0$,
\begin{equation}
\mathrm{s-}\!\lim_{t\to 0} U_{\mathrm{Z}}(t) =P .
\end{equation}
Then there is a semibounded self-adjoint operator $H_Z$ such that
$H_{\mathrm{Z}}=P H_{\mathrm{Z}} P$ and
\begin{equation}
U_{\mathrm{Z}}(t) = P \exp(-i H_{\mathrm{Z}} t)
\label{eq:MS}
\end{equation}
for all $t\in\mathbb{R}$. Moreover, $H_{\mathrm{Z}}|_{\cH_P}$ is uniquely
associated with the closed and densely defined quadratic form
$\mathfrak{h}_{\mathrm{Z}} : \mathcal{H}_P \cap D(H^{1/2})\to
\mathbb{R}$
\begin{equation}
\mathfrak{h}_{\mathrm{Z}}(\varphi)= \| H^{1/2} \varphi \|^2 ,
 \end{equation}
that is,
\begin{equation}
\label{eq:MSHZ}
H_{\mathrm{Z}} = (H^{1/2} P)^\dagger (H^{1/2} P).
 \end{equation}

\subsection{Remarks}

The consequence of the theorem is straightforward. By Eq.\ (\ref{eq:MS}) the density matrix after the Zeno evolution (\ref{eq:Nproie}) is
\andy{infproieaaa}
\beq
\rho (t) = \lim_{N\to\infty} V_N(t) \rho_0 V_N(t)/p^{(N)}(t) = U_{\mathrm{Z}}(t)  \rho_0 U_{\mathrm{Z}}^\dagger (t)
  \label{eq:infproieaaa}
\eeq
and the probability to find the system in ${\cal H}_P$ at the final
time $t$ is
\andy{probinfobaaa}
\beq
p(t) =\lim_{N \rightarrow \infty} p^{(N)}(t)
   = \mbox{Tr} \left[ U_{\mathrm{Z}}(t)  \rho_0 U_{\mathrm{Z}}^\dagger (t) \right]
 = \mbox{Tr} \left[ \rho_0 P \right] = 1 .
\label{eq:probinfu}
\eeq
If the particle is ``continuously" observed, in order to check
whether it has survived inside ${\cal H}_P$, it will never make a
transition to  ${\cal H}_P^\perp$. This is the original formulation
of the quantum Zeno paradox.

Note that the continuity condition at $t=0$ is equivalent to
requiring that $D(H^{1/2})$ be dense in $\cH_P$. Therefore, the
proof is the combination of Kato's product formula
(\ref{eq:ZenoKato}), which is valid for self-adjoint semigroups,
with an analytic continuation. The latter part relies on the
following technical lemma, whose proof can be found in
\cite{Misra77}. Here we will follow the modified proof by Exner
\cite{Exner}, who also gives the explicit expression of the Zeno
Hamiltonian $H_{\mathrm{Z}}$. See also \cite{Schmidt04}.

We conclude by remarking that Kato's product formula, that greatly
simplifies the proof, appeared in 1978 \cite{Kato1}, one year after
the article by Misra and Sudarshan \cite{Misra77}. The two
publications were independent: the first focused on functional
analysis, the second on mathematical physics.

\subsection{Lemma}
For each $N$, the function $V_N(z)$ is defined and strongly
continuous in the closed lower halfplane $\{z\in\mathbb{C}, \Im z
\leq 0\}$ and is strongly analytic in the open lower halfplane. The
following integral relations hold \cite{Misra77}
\begin{equation}
\label{eq:Cauchy1}
V_N(z) = \frac{(z-i)^2}{2\pi i} \int_{\mathbb{R}}
\frac{V_N(t)}{(t-i)^2 (z-t)}\, dt, \qquad \Im z<0,
\end{equation}
\begin{equation}
\label{eq:Cauchy2}
 \frac{1}{2\pi i} \int_{\mathbb{R}} \frac{V_N(t)}{(t-i)^2 (z-t)}\, \de t =0, \qquad \Im z>0.
\end{equation}

\subsection{Proof of the theorem}

We start from (\ref{eq:Cauchy1}). The limit
$\mathrm{s-}\lim_{N\to\infty}  V_N(t)$ exists by assumption for all
$t\in\mathbb{R}$, and by dominated convergence one gets
\begin{equation}
U_{\mathrm{Z}}(z)= \frac{(z-i)^2}{2\pi i} \int_{\mathbb{R}}
\frac{U_{\mathrm{Z}}(t)}{(t-i)^2 (z-t)}\, \de t, \qquad \Im z<0.
\end{equation}
Now for each $t>0$, by Kato's product formula
\begin{equation}
U_{\mathrm{Z}}(-i t) =\mathrm{s-}\!\!\!\lim_{N\to\infty}  (P \exp(-H
t/N) P)^N = P \exp(-H_{\mathrm{Z}} t).
\label{eq:halfline}
\end{equation}
It is easy to see that $z\mapsto P \exp(-i H_{\mathrm{Z}} z)$ is
strongly analytic for all $z$ with $\Im z<0$. Now, the two
operator-valued analytic functions $P \exp(-i H_{\mathrm{Z}} z)$ and
$U_{\mathrm{Z}}(z)$ coincide on the half line $z=-i t$, with $t \in
\mathbb{R}_+$, see Eq.\ (\ref{eq:halfline}), and thus they coincide
on the whole half plane
\begin{equation}
\label{eq:Zenohalfplane}
U_{\mathrm{Z}}(z) = P \exp(-i H_{\mathrm{Z}} z), \qquad
\Im z <0.
\end{equation}
Moreover, by functional calculus one gets the group relation
$U_{\mathrm{Z}}(z_1)U_{\mathrm{Z}}(z_2)=U_{\mathrm{Z}}(z_1+z_2)$ for
all $z$ with $\Im z <0$.

Finally, it is not difficult to show \cite{Misra77} that
$U_{\mathrm{Z}}(t)$ is the boundary value of the above analytic
function, namely,
\begin{equation}
\mathrm{s-}\!\lim_{\varepsilon\downarrow 0} U_{\mathrm{Z}}(t-i\varepsilon) = U_{\mathrm{Z}}(t)
\label{eq:boundaryvalue}
\end{equation}
holds for all $t\in\mathbb{R}$. By combining
(\ref{eq:Zenohalfplane}) with (\ref{eq:boundaryvalue}) one gets the
desired result (\ref{eq:MS}).

\section{Existence of the limit}
\label{sec-EI}

In the theorem by Misra and Sudarshan the existence of the Zeno
limit is postulated. Clearly, it remains to prove that the limit
exists  for $H$ unbounded and $P$ infinite dimensional. Once the
limit is proven to exist, it must have the form (\ref{eq:MSHZ}).
Many efforts have been done in this direction during the last two or
three decades.

In 2004, Exner and Ichinose proved the existence of the limit in a
weak sense
\cite{ExnerIchinose}. The convergence is only in
$L^2_{\mathrm{loc}}(\mathbb{R}, \cH)=\{\varphi: \mathbb{R} \to \cH |
\int_K \|\varphi(t) \|^2 \de t < \infty, \mbox{ for every compact } K \subset\mathbb{R}
 \}$.
The statement of the theorem is the following: if $H\geq 0$ and
$D(H^{1/2}P)$ is dense in $\cH$, then for any $\varphi\in\cH$
\begin{equation}
\lim_{N\to\infty} \int_0^{s} \left\| V_N(t) \varphi - P \exp(-i H_{\mathrm{Z}} t) \varphi \right\|^2 \de t =0,
\label{eq:mediat}
\end{equation}
for any $s>0$, with
\begin{equation}
H_{\mathrm{Z}}=(H^{1/2} P)^\dagger (H^{1/2} P).
\end{equation}
This, in turn, yields the existence of the limit of the Zeno product formula
for almost all $t$ in the strong operator topology
along a suitable increasing subsequence
$\{N_j\}_{j\in\mathbb{N}}\subset \mathbb{N}$ of natural numbers:
\begin{equation}
\mathrm{s-}\!\lim_{j\to \infty} V_{N_j}(t) = P \exp(-i H_{\mathrm{Z}} t), \quad \mbox{for a.e. }
t\in\mathbb{R}.
\end{equation}
This is  the state of the art in the Zeno product formula. In our
opinion, it is a satisfactory result from a physical standpoint.
From a mathematical perspective, however, one might still hope to
prove a  stronger result.

\section{Corollary: Position measurements}
\label{sec-coroll}

Let us conclude our mathematical discussion with a particular case
of physical interest: the position measurement of a particle in a
well-behaved potential. See Sec.\
\ref{sec-infinitedim} and in particular
\ref{sec-potentialzeno}.

Let $H= -\frac{1}{2 m} \Delta + V(x)$ in $L^2(\mathbb{R}^d)$, $V\in
L^\infty (\mathbb{R}^d)$ bounded, and $P=\chi_\Omega(x)$ with
$\Omega\subset\mathbb{R}^d$ an open set with regular boundary
$\partial\Omega$.  The Zeno limit exists in the
$L^2_{\mathrm{loc}}(\mathbb{R}, \cH)$ topology and is of the form
(\ref{eq:MS}) with
\begin{equation}
H_{\mathrm{Z}} = \left(-\frac{1}{2 m} \Delta_{\Omega} + V(x)\right)
\chi_{\Omega}(x),
\label{eq:Zeno potential}
\end{equation}
where $\Delta_{\Omega}$ is the Dirichlet Laplacian on
$\cH_P=L^2(\Omega)$.

This is the rigorous statement behind the physical proof of Sec.\
\ref{sec-infinitedim}. Assume that one frequently checks whether a
$d$-dimensional quantum system (particle) is contained in a spatial
region $\Omega$. The Zeno effect takes place and the dynamics is
governed by the Hamiltonian (\ref{eq:Zeno potential}). The
convergence in the $L^2_{\mathrm{loc}}(\mathbb{R}, \cH)$ topology,
rather than in the strong topology, is tantamount to assuming a time
coarse graining over a small time interval $s$: see
(\ref{eq:mediat}).

\subsection{Proof}
$H$ is self-adjoint and semibounded, since it is a bounded
perturbation of the Laplacian. Without loss of generality we can
assume $V(x)\geq 0$, whence $H\geq 0$. Obviously, the set of smooth
functions of compact support contained in  $\Omega$ satisfies
$C^{\infty}_0(\Omega)\subset D(H^{1/2})\cap \cH_P$ and it is well
known to be dense, $\overline{C^{\infty}_0(\Omega)}=\cH_P$. Thus
$\overline{D(H^{1/2}P)}=\cH$ and the theorem applies. The
restriction of the Zeno Hamiltonian $H_{\mathrm{Z}}|_{\cH_P}$ is
associated with the closure of the quadratic form
\begin{equation}
\varphi \in C^{\infty}_0(\Omega) \mapsto \|H^{1/2} \varphi\|^2 =
\frac{1}{2} \int_\Omega | \nabla \varphi(x) |^2 \de^d x +
\int_\Omega V(x) |\varphi(x)|^2 \de^d x .
\end{equation}
However, due to the boundness of $V$, the domain of the closure is
nothing but $\{\varphi \in L^2(\Omega) | \nabla \varphi \in
L^2(\Omega), \; \varphi(\partial \Omega)=0\}= H^1_0(\Omega)$. The
vectors $\varphi\in H_0^1(\Omega)$ in the domain of
$H_{\mathrm{Z}}|_{\cH_P}$ should in addition satisfy $H\varphi \in
\cH_P$, and, due to the boundness of $V$, this implies
 that $\varphi \in H^2(\Omega)=\{\varphi \in L^2(\Omega) | \Delta \varphi \in
L^2(\Omega)\}$. Therefore $D(H_{\mathrm{Z}})\cap \cH_P=
H^2(\Omega)\cap H^1_0(\Omega)$,  which is the domain of the
Dirichlet Hamiltonian $\Delta_\Omega$, and the desired result is
obtained.

\section{Three alternative ways to obtain the Zeno subspaces}
\label{sec-zenosubsp}

After the mathematical interlude of Secs.\
\ref{sec-bounded}-\ref{sec-coroll}, we revert to a
less rigorous analysis and focus on applications. The quantum Zeno
phenomenon is usually ascribed to repeated von Neumann's
projections on a quantum system. Indeed, this is the approach we
have adopted so far. In a way, this approach goes back to Misra
and Sudarshan \cite{Misra77} and to some extent, even to von
Neumann \cite{vonNeumann32}.

However, during the last few years it has become clear that this
view of the QZE is too narrow, because the projective measurements
can be replaced by another quantum system interacting strongly with
the principal system. The QZE appears therefore to be a more general
phenomenon, that can be explained in dynamical terms. After all, a
projection \textit{\`a la} von Neumann is just a handy way to
summarize the complicated physical processes that take place during
a quantum measurement. The latter is performed by an external
apparatus or a quantum field and may involve complicated
interactions with the environment. The external system performing
the observation need not be a \textit{bona fide} detection system,
that clicks or is endowed with a pointer. It is enough that the
information on the state of the observed system be encoded in some
external degrees of freedom by a physical process that associates
different (external) states to different values of the observable
being measured. For instance, a spontaneous emission process can be
a very effective measurement, for it is irreversible and entangles
the state of the system (the emitting atom or molecule) with the
state of the apparatus (the electromagnetic field). The von Neumann
rules arise when one traces away the photonic state and is left with
an incoherent superposition of atomic states. In the light of these
observations, it is clear that the main physical features of the
Zeno effect are a consequence of the dynamics and need not be
ascribed to the ``collapse" of the wave function. But then, one
would like to understand which features of the dynamical process are
essential for observing a QZE. It turns out that the QZE takes place
whenever a strong disturbance ``dominates" the time evolution of the
quantum system.

It is worth emphasizing that it is not only physically reasonable,
but also logically appealing to view the QZE as a dynamical
effect: in this broader context, different decoupling and control
schemes can be understood as arising from the same physical
considerations, and hence can be unified under the same conceptual
and formal framework. Furthermore, they appear as particular cases
of a more general dynamics in which the system of interest is
strongly coupled to an external system that (loosely speaking)
plays the role of a measuring apparatus.

We now discuss three different manifestations of the quantum Zeno
effect. We start in Sec.\ \ref{sec-nonselect} with (projective)
measurements, then extend the notion of QZE to the case of unitary
kicks in Sec.\ \ref{sec-qmaps} and finally discuss (unitary)
continuous interactions in Sec.\ \ref{sec-contQZE}. In extending
the notion of QZE to unitary processes we shall also find it
convenient to study the evolution in the whole Hilbert space, that
will be split into invariant, Zeno subspaces. In the two latter
cases (unitary kicks and continuous coupling) the quantum Zeno
subspaces will turn out to be the eigenspaces of the interaction.
We shall discuss the superselection rule that originates from the
Zeno dynamics in Sec.\ \ref{sec-dynssQZE} and show the close
equivalence between the two unitary approaches in Sec.\
\ref{sec-unitequiv}.

\subsection{Quantum Zeno subspaces via projective measurements}
\label{sec-nonselect}

We first consider projective von Neumann's measurements. Besides
being incomplete, in the sense specified at the beginning of Sec.\
\ref{sec-finitedim}, the quantum measurements will be
``nonselective," in the sense that the measuring apparatus does
not select the different outcomes, but simply destroys the phase
correlations between some states, provoking the transition from a
pure state to a mixture. See, for example,
\cite{Schwinger59,Peres98}.

We now extend Misra and Sudarshan's theorem \cite{Misra77} to
incomplete and nonselective measurements \cite{subspaces}. Let the
evolution of the quantum system be described by the superoperator
\beq
\hat U_t \rho=U(t) \rho U^\dagger(t),\qquad U(t)=\exp(-i H t)
\eeq
where $\rho$ is the density matrix of the system and $H$ a
time-independent lower-bounded Hamiltonian. Let
\beq
\{P_n\}_n, \qquad
P_nP_m=\delta_{mn}P_n,\qquad  \sum_n P_n=1 ,
\eeq
be a finite orthogonal resolution of the identity and $P_n\cH=\cH_{n}$
the relative subspaces. The Hilbert space is accordingly
partitioned in
\beq
\label{eq:partition}
\cH=\bigoplus_n \cH_{n}.
\eeq
The nonselective measurement is described by the superoperator
\beq
\label{eq:superP} \hat P \rho=\sum_n P_n \rho P_n
\eeq
and the evolution after $N$ measurements in a time $t$ is governed
by the superoperator
\beq
\label{eq:Nevol}
\hat V^{(N)}_t=\underbrace{\left(\hat P\hat U_{t/N}\right)
\left(\hat P\hat U_{t/N}\right)\cdots \left(\hat P\hat
U_{t/N}\right)}_{N\; \mathrm{times}} =\left(\hat P \hat
U_{t/N}\right)^{N} .
\eeq
Let us prepare the system in the initial state
\beq
\label{eq:inrho}
\hat P \rho_0=\sum_n P_n \rho_0 P_n.
\eeq
The evolution reads
\beq
\rho(t)=\hat V^{(N)}_t \rho_0 =\sum_{n_1,\dots,n_N}V_{n_1\dots
n_N}^{(N)}(t)\; \rho_0\; V_{n_1\dots n_N}^{(N)\dagger}(t) ,
\eeq
where
\barr
V_{n_1\dots n_N}^{(N)}(t)  = P_{n_N} U\left(t/N\right) P_{n_{N-1}}
\cdots P_{n_2} U\left(t/N\right) P_{n_1}, \label{eq:boo}
\earr
that should be compared to Eq.\ (\ref{eq:Nproie}) (which is obtained
as a particular case when all projectors are the same). We assume,
like in Sec.\
\ref{sec-MStheorem}, the existence of
the strong limits
\barr
& & U_{\rm Z}^{(n)} (t)=\mathrm{s-}\!\lim_{N\to\infty} V_{n\dots
n}^{(N)}(t) =\lim_{N\to\infty} \left[P_n U\left(\frac{t}{N}\right)
P_n \right]^N ,
\\
& & \mathrm{s-}\!\lim_{t \rightarrow 0^+} U_{\rm Z}^{(n)} (t)= P_n ,
\quad \forall n \
\label{eq:slims}
\earr
Then $U_{\rm Z}^{(n)}(t)$ form a semigroup, and
\beq
U_{\rm Z}^{(n)\dagger}(t)U_{\rm Z}^{(n)}(t) =P_n.
\eeq
Moreover, it is easy to show that
\andy{fulldiag}
\beq
\lim_{N\to\infty} V_{n\dots n'\dots}^{(N)}(t) = 0, \qquad
\mathrm{for}\quad n'\neq n .
\label{eq:fulldiag}
\eeq
Notice that, for any \emph{finite} $N$, the off-diagonal operators
(\ref{eq:boo}) are in general nonvanishing, i.e.\ $V_{n\dots
n'\dots}^{(N)}(t) \neq 0$ for $n'\neq n$. It is only in the limit
(\ref{eq:fulldiag}) that these operators become diagonal. This is
because $U\left(t/N\right)$ provokes transitions among different
subspaces $\cH_n$. The limiting evolution superoperator is
\beq
\label{eq:limseqV}
\hat U_{\rm Z} (t) \equiv \lim_{N\to\infty} \hat V^{(N)}_t,
\eeq
and the final state reads
\andy{rhoZ}
\barr
\rho(t) & = & \hat U_{\rm Z} (t) \rho_0 =\sum_n U_{\rm Z}^{(n)} (t)
\rho_0 U_{\rm Z}^{(n)\dagger} (t),
\nonumber \\
& &  \mathrm{with} \quad \sum_n U_{\rm Z}^{(n)\dagger}(t) U_{\rm
Z}^{(n)} (t) =\sum_n P_n=\mathbf{1} .\;\; \label{eq:rhoZ}
\earr
The components $U_{\rm Z}^{(n)} (t)
\rho_0 U_{\rm Z}^{(n)\dagger} (t)$ make up a block diagonal
matrix: the initial density matrix is reduced to a mixture and any
interference between different subspaces $\cH_{n}$ is destroyed
(complete decoherence). Moreover,
\barr
\fl\qquad p_n(t) =  \mathrm{Tr} \left[\rho(t)
P_n\right]=\mathrm{Tr}\left[U_{\rm Z}^{(n)} (t)
\rho_0 U_{\rm Z}^{(n)\dagger} (t)\right]= \mathrm{Tr}\left[\rho_0 P_n\right]=p_n(0)
, \quad \forall n .
\label{eq:probinfuaaa}
\earr
Probability is conserved in each subspace and no probability
leakage between different subspaces is possible: the total
Hilbert space splits into invariant \emph{Zeno
subspaces}\index{quantum Zeno subspaces} $\cH_n$ and the different
components of the density matrix independently evolve within each
sector. One can think of the total Hilbert space as the shell of a
tortoise, each invariant subspace being one of the scutes. Motion
among different scutes is impossible. (See Fig.\
\ref{tortoise1} in the following.) Misra and Sudarshan's
seminal result is reobtained when $p_n(0)=1$ for some $n$, in
(\ref{eq:probinfuaaa}): the initial state is then in one of the
invariant subspaces and the survival probability in that subspace
remains unity.

When the Hamiltonian is bounded $\| H\|<\infty$, each limiting
evolution operator $U_{\rm Z}^{(n)}$ in (\ref{eq:slims}) is
unitary within the subspace $\cH_{n}$ and has the form
\andy{cVfin}
\beq
U_{\rm Z}^{(n)}(t)= \lim_{N\to\infty} [ P_n U(t/N) P_n]^N = P_n
\exp(-i P_n H P_n t)  \label{eq:cVfin} .
\eeq
More generally, if $\cH_{n}\subset D (H)$ (which is trivially
satisfied for a bounded $H$), then the resulting Hamiltonian $P_n H
P_n$ is self-adjoint and $U_{\rm Z}^{(n)}(t)$ is unitary in
$\cH_{n}$. When the above condition does not hold, one has to resort
to the theorem proved in Sec.\ \ref{sec-MStheorem} and work out the
real form of the self-adjoint Zeno Hamiltonian $H_{\rm Z}^{(n)}=P_n
H_{\rm Z}^{(n)} P_n$ in the sector $\cH_n$.

In any case, with the necessary precautions on the meaning of
operators and boundary conditions, the Zeno evolution can be
written
\beq
\hat U_{\rm Z} (t) \rho_0 =\sum_n P_n \exp(-i H_{\mathrm{Z}}t )
\rho_0 \exp(i H_{\mathrm{Z}}t ) P_n,
\eeq
where
\beq
H_{\mathrm{Z}}=\hat P H =\sum_n P_n H P_n
\label{eq:ZenoHam}
\eeq
is the global Zeno Hamiltonian.

\subsection{Quantum Zeno subspaces via unitary kicks (``bang-bang")}
\label{sec-qmaps}
\andy{sec-qmaps}

We have seen that if the projections are multidimensional, the
system evolves in a collection of Zeno subspaces. The \emph{deus
ex machina} of these phenomena are von Neumann's projections, that
are supposed to be
\emph{instantaneous} processes, yielding the collapse of the wave
function (an ultimately nonunitary process). However, QZE is
\emph{not} a consequence of nonunitary evolutions: it can be
obtained by repeatedly dividing the wave function into branch waves
\cite{Pascazio94,Pascazio97} (but see also \cite{Petrosky90}). If the branching
processes are frequent enough, one gets again Zeno. We now further
elaborate on this issue, obtaining first, in this subsection, the
quantum Zeno subspaces by means of a sequence of frequent
instantaneous unitary processes, then in the next subsection by
means of a strong continuous coupling. We will only sketch the main
results: additional details and a complete proof can be found in
\cite{FLP,trieste}.

Consider the dynamics of a quantum system  undergoing $N$
``kicks" $U_{\mathrm{kick}}$ in a time interval $t$. Kicks are
simply instantaneous unitary transformations, in practice a
limiting concept (the duration of the kick being the shortest
timescale in the problem at hand). Notice the similarity with a
von Neumann projection, a process that is also supposed to take
place instantaneously. Consider a system that undergoes a smooth
unitary evolution $U$ interspersed at equal time intervals $t/N$
with $N$ kicks. The evolution reads
\barr
U_N(t) &=&
\underbrace{\left[U_{\mathrm{kick}}U\left(\frac{t}{N}\right)
\right]\left[U_{\mathrm{kick}}U\left(\frac{t}{N}\right)
\right]\cdots \left[U_{\mathrm{kick}}U\left(\frac{t}{N}\right)
\right]}_{N\; \mathrm{times}} \nonumber \\
& = &
\left[U_{\mathrm{kick}}U\left(\frac{t}{N}\right)
\right]^N
\label{eq:BBevol}
\earr
In the large $N$ limit, the evolution is dominated by the large
contribution of $(U_{\mathrm{kick}})^N$. One therefore considers
the sequence $\{V_N \}_N$ of unitary operators
\andy{eq:sequence}
\barr
\label{eq:sequence}
V_N(t)&=&(U_{\mathrm{kick}}^{\dagger})^N U_N(t)=
(U_{\mathrm{kick}}^{\dagger})^N \left[U_{\mathrm{kick}}
U\left(\frac{t}{N}\right) \right]^N
\earr
and its limit
\andy{eq:limseq}
\beq
\label{eq:limseq}
U_{\rm Z}(t) \equiv \lim_{N\to\infty} V_N(t).
\eeq
One can show that
\andy{eq:eqUz}
\beq
\label{eq:eqUz}
U_{\rm Z}(t) = \exp(-i H_{\mathrm{Z}} t),
\eeq
where
\andy{eq:eqHz}
\beq
\label{eq:eqHz}
H_{\mathrm{Z}} = \hat P H = \sum_n P_n H P_n
\eeq
is the Zeno Hamiltonian, $P_n$ being the spectral projections of
$U_{\mathrm{kick}}$
\andy{eq:specdec}
\beq
U_{\mathrm{kick}}=\sum_n \mathrm{e}^{-i\lambda_n} P_n . \quad
(\mathrm{e}^{-i\lambda_n}\neq \mathrm{e}^{-i\lambda_l}, \; \mbox{for} \; n\neq l),
\label{eq:specdec}
\eeq
that we assume to have a discrete spectrum. In conclusion
\andy{eq:evolNZ}
\barr
U_N(t) &\sim & U_{\mathrm{kick}}^N U_{\rm Z}(t)=
U_{\mathrm{kick}}^N
\exp(-i H_{\mathrm{Z}} t) \nonumber \\
&=&\exp\left(-i \sum_n  N \lambda_n P_n + P_n H P_n t \right).
\label{eq:evolNZ}
\earr
This is again a Zeno dynamics, yielding Zeno subspaces, the
partition of the Hilbert space depending now on the features of
the kick operator (\ref{eq:specdec}). The situation is identical
to the case of repeated projective measurements discussed in Sec.\
\ref{sec-nonselect}.

It is remarkable to observe that in this case the map
$H_{\mathrm{Z}}=\hat P H$ is the projection onto the centralizer
\andy{eq:centro}
\beq
\label{eq:centro}
Z(U_{\mathrm{kick}})=\{X|\; [X, U_{\mathrm{kick}}]=0\}.
\eeq
The appearance of the Zeno subspaces is a direct consequence of
the wildly oscillating phases between different eigenspaces of the
kick (yielding a superselection rule \cite{WWW,WWW1}) and hinges
on von Neumann's ergodic theorem \cite{ReedSimon}.

The analogy of the approach outlined in this section with the
seminal papers on quantum maps and quantum chaos \cite{qch,qch2} is
manifest. Note, however, that here we are interested in the limit
$\tau=t/N\to0$, with $t$ finite, while in quantum chaos the main
interest is  in the large time limit $t\to\infty$, with $\tau$
finite. The efficacy of ``bang-bang" kicks in controlling the
dynamics in NMR experiments is well known since the sixties
\cite{anderson,ernst,freeman,levitt} and was
revived thirty years later in the context of quantum information
\cite{viola98}. An excellent review of these techniques can be found in
\cite{lidarrev}.

\subsection{Quantum Zeno subspaces via a strong continuous coupling}
 \label{sec-contQZE}
 \andy{contQZE}

Both von Neumann's projections and unitary kicks are limiting
processes, that are supposed to take place instantaneously, namely
on a very short timescale when compared to the other timescales
characterizing the evolution of the quantum system. On the other
hand, short timescales can be physically associated with strong
couplings. It is then natural to expect that the essential features
of the QZE can be obtained by making use of a strong continuous
coupling, when the external system takes a sort of steady, powerful
``gaze" at the system of interest. The mathematical formulation of
this idea is contained in a theorem on the (large-$K$) dynamical
evolution governed by a generic Hamiltonian of the type
\andy{HKcoup}
\beq
H_K= H + K H_{\mathrm{c}} ,
 \label{eq:HKcoup}
\eeq
where $H$ is the Hamiltonian of the quantum system, $H_{\mathrm{c}}$
an additional interaction Hamiltonian caricaturing the ``continuous
measurement" and $K$ a coupling constant.

In the limit $K\to\infty$ (``infinitely strong measurement" or ``infinitely quick
detector"), the evolution operator
\barr
U_{K}(t) = \exp(-iH_K t)  \label{eq:measinter}
\earr
is dominated by $\exp(-i K H_{\mathrm{c}}t)$. One therefore
considers the limiting operator
\beq
\label{eq:limevol}
U_{\rm Z}(t) = \lim_{K\to\infty}\exp(i K
H_{\mathrm{c}}t)\,U_{K}(t),
\eeq
that can be shown to have the form
\beq
\label{eq:theorem}
U_{\rm Z}(t)=\exp(-i H_{\mathrm{Z}} t),
\eeq
where
\beq
H_{\mathrm{Z}}=\hat P H =\sum_n P_n H P_n \label{eq:diagsys}
\eeq
is the Zeno Hamiltonian, $P_n$ being the eigenprojection of
$H_{\mathrm{c}}$, that we suppose to have a discrete spectrum,
belonging to the eigenvalue $\eta_n$
\beq
\label{eq:diagevol}
H_{\mathrm{c}} = \sum_n \eta_n P_n, \qquad (\eta_n\neq\eta_m,
\quad \mbox{for} \; n\neq m) \ .
\eeq
This is formally identical to (\ref{eq:ZenoHam}) and
(\ref{eq:eqHz}). In conclusion, the limiting evolution operator is
\andy{eq:measinterbis}
\barr
U_K(t)\sim\exp(-i K H_{\mathrm{c}}t)\,U_{\rm Z}(t) =
\exp\left(-i\sum_n K t\eta_n P_n + P_n H P_n t\right) ,
\label{eq:measinterbis}
\earr
whose block-diagonal structure is explicit and yields the Zeno
subspaces. Compare with (\ref{eq:evolNZ}). The above statements can
be proved by making use of the adiabatic theorem. Like in the
previous subsections, where the Zeno dynamics was obtained by making
use of frequent kicks, $\hat P$ in (\ref{eq:diagsys}) projects onto
the centralizer
\beq
\label{eq:centro2}
Z(H_{\mathrm{c}})=\{X|\; [X, H_{\mathrm{c}})]=0\}.
\eeq
Again, the Zeno subspaces are a consequence of the wildly
oscillating phases between different eigenspaces.

The notion of a continuous observation of the quantum state,
performed for example by its environment or an intense field, dates
back to the eighties. Chiral molecules can exist in two
reflection-related isomers, but in practice they only appear as one
or the other isomer and never in their symmetric superposition (the
system ground state). Simonius \cite{simonius} and then Harris and
Stodolski \cite{HS} argued that the solution containing the
molecules acts as an environment that continuously observes the
molecules, decohering them and inhibiting any transitions. This
concept is similar, in embryo, to that discussed in this subsection.
Similar ideas were discussed in literature of the last two decades
\cite{Peres80,Kraus81,sudbery,Venugopalan95,Plenio96,Berry96,Mihokova97,luissanchez,Thun98,panov,Rehacek,Facchi00,MMN,luis}.
The first quantitative estimate of the link with the formulation in
terms of projective measurements is rather recent
\cite{Mihokova97,lss98,rev3}.

\subsection{Dynamical superselection rules}
\label{sec-dynssQZE}
\andy{dynssQZE}

\begin{figure}
\begin{center}
\includegraphics[height=6.5cm]{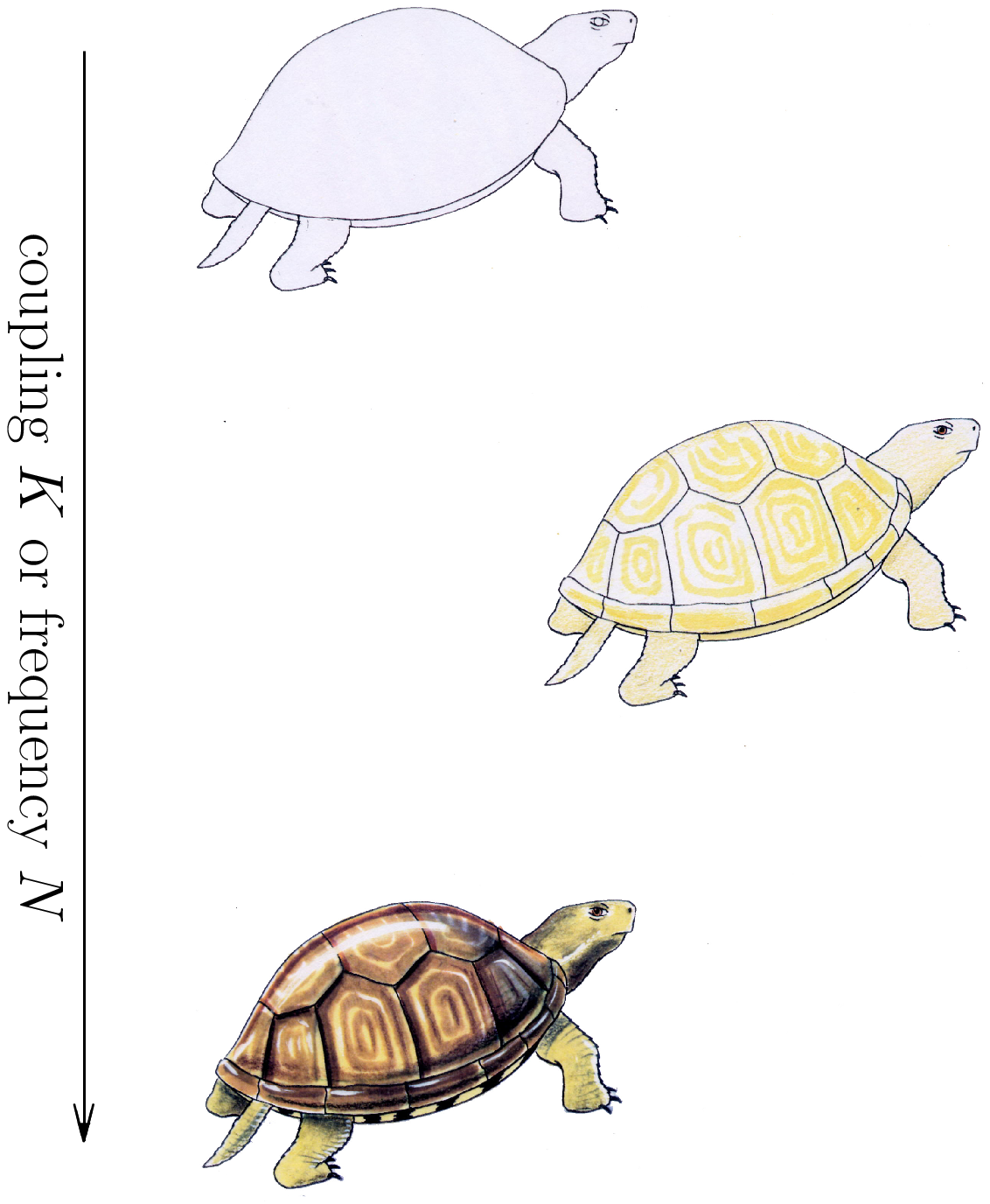}
\end{center}
\caption{The Zeno subspaces. The Hilbert space of the system splits into
sectors (the scutes of the shell of a tortoise) and a dynamical
superselection rule appears as the number of measurements/kicks $N$
or the coupling $K$ is increased. This drawing blends the paradox of
the sped arrow that never reaches its target with that of the
tortoise and swifter-running Achilles, which was also proposed by
Zeno in order to counter the idea of motion: in a race in which the
tortoise has a head start, Achilles can never overtake it, because
before he comes up to the point at which the tortoise started, the
tortoise will have got a little way, and so on ad infinitum: ``that
which is in locomotion must arrive at the half-way stage before it
arrives at the goal"
\cite{aristotle}. In our quantum context, sped arrows can only move
\emph{within} the scutes, but never cross the boundary between
different scutes.}
\label{tortoise1}
\end{figure}

Let us briefly discuss the physics behind the different
manifestations of the quantum Zeno effect discussed in this section.
In the $N\to \infty$ ($K\to\infty$) limit the time evolution
operator $U_{\rm Z}(t)$ becomes diagonal with respect to
$U_{\mathrm{kick}}$ or $H_{\mathrm{c}}$, i.e.\ it belongs to their
centralizers,
\beq
[U_{\rm Z} (t), U_{\mathrm{kick}}]=0,\qquad [U_{\rm Z} (t),
H_{\mathrm{c}}]=0,
\eeq
a superselection rule arises and the total Hilbert space is split
into subspaces $\cH_{n}$ that are invariant under the evolution.
The dynamics within each Zeno subspace $\cH_{n}$ is governed by
the Zeno Hamiltonian $H_{\mathrm{Z}}P_n =P_n H P_n$, which is the
diagonal part of the system Hamiltonian $H$, the remaining part of
the evolution consisting in a sector-dependent phase. The
probability to find the system in each $\cH_{n}$
\barr
p_n(t)&=&\mathrm{Tr} \left[ \rho(t) P_n \right]= \mathrm{Tr}
\left[U_{\rm Z} (t)\rho_0U_{\rm Z}^\dagger (t) P_n\right] =\mathrm{Tr}
\left[U_{\rm Z}(t)\rho_0 P_nU_{\rm Z}^\dagger(t)\right]
\nonumber\\
&=& \mathrm{Tr} \left[ \rho_0 P_n \right]=p_n(0)
\label{eq:pntpn0}
\earr
is constant. As a consequence, if the initial state is an incoherent
superposition of the form (\ref{eq:inrho}), then each component will
evolve separately, according to
\beq
\rho(t)=U_{\rm Z}(t)\rho_0U_{\rm Z}^\dagger(t)= \sum_n U_{\rm Z}^{(n)}(t) \rho_0
U_{\rm Z}^{(n)\dagger}(t),
\label{eq:cUcV}
\eeq
with $U_{\rm Z}^{(n)}(t)= P_n \exp(-i P_n H P_n t)$, which is
exactly the same result (\ref{eq:rhoZ})-(\ref{eq:cVfin}) found in
the case of projective measurements. In Fig.\ \ref{tortoise1} we
endeavored to give a pictorial representation of the decomposition
of the Hilbert space in the three cases discussed (projective
measurements, kicks and continuous coupling).

Notice, however, that there is one important difference between
the nonunitary evolution discussed in Sec.\ \ref{sec-nonselect}
and the dynamical evolutions discussed in Secs.\
\ref{sec-qmaps}-\ref{sec-contQZE}: indeed, if the initial state
$\rho_0$ contains coherent terms between any two Zeno subspaces
$\cH_{n}$ and $\cH_{m}$, $P_n \rho_0 P_m\neq 0$, these vanish
after the first projection (\ref{eq:rhoZ}) in Sec.\
\ref{sec-nonselect}: $P_n \rho(0^+) P_m = 0$ [the state becomes an
incoherent superposition $\rho(0^+)\neq\rho_0$, whence $\mathrm{Tr}
\rho(0^+)^2 < \mathrm{Tr} \rho_0^2$]. On the other hand, such terms
are preserved by the dynamical (unitary) evolutions analyzed in
Secs.\ \ref{sec-qmaps}-\ref{sec-contQZE}, and do not vanish, even
though they wildly oscillate. For example, consider the initial
state
\beq
\rho_0= (P_n+P_m)\rho_0 (P_n+P_m), \qquad P_n \rho_0 P_m \neq 0.
\eeq
By (\ref{eq:evolNZ}) and (\ref{eq:measinterbis}) it evolves into
\barr
\rho(t)&=&U_{\rm Z}^{(n)}(t) \rho_0 U_{\rm Z}^{(n)\dagger}(t)+ U_{\rm Z}^{(m)} (t) \rho_0
U_{\rm Z}^{(m)} (t)
\nonumber\\
& & + \; \mathrm{e}^{-i N (\lambda_n-\lambda_m)} U_{\rm Z}^{(n)} (t) \rho_0
U_{\rm Z}^{(m)\dagger}(t)
\nonumber\\
& & + \; \mathrm{e}^{i N (\lambda_n-\lambda_m)} U_{\rm Z}^{(m)} (t) \rho_0
U_{\rm Z}^{(n)\dagger} (t)
\earr
or
\barr
\rho(t)&=&U_{\rm Z}^{(n)}(t) \rho_0 U_{\rm Z}^{(n)\dagger}(t)+ U_{\rm Z}^{(m)} (t) \rho_0
U_{\rm Z}^{(m)} (t)
\nonumber\\
& & + \; \mathrm{e}^{-i K (\eta_n-\eta_m) t} U_{\rm Z}^{(n)} (t) \rho_0
U_{\rm Z}^{(m)\dagger}(t)
\nonumber\\
& & + \; \mathrm{e}^{i K (\eta_n-\eta_m) t} U_{\rm Z}^{(m)} (t) \rho_0
U_{\rm Z}^{(n)\dagger} (t) ,
\earr
respectively, at variance with (\ref{eq:rhoZ}). Therefore
$\mathrm{Tr} \rho(t)^2 =
\mathrm{Tr} \rho_0^2$ for any $t$ and the Zeno dynamics is unitary
in the \textit{whole} Hilbert space $\cH$. We notice that these
coherent terms become unobservable in the large-$N$ or large-$K$
limit, as a consequence of the Riemann-Lebesgue theorem (applied to
any observable that ``connects" different sectors and whose time
resolution is finite). This interesting aspect is reminiscent of
some results on ``classical" observables \cite{Jauch}, semiclassical
limit \cite{Berry} and quantum measurement theory
\cite{Schwinger59,araki1,machidanamiki1,machidanamiki2,araki2}.
It is also interesting to note that the superselection
rules\index{superselection rules} discussed here are
\textit{de facto} equivalent to the celebrated ``W$^3$" ones
\cite{WWW,WWW1}, but turn out to be a mere consequence of the Zeno
dynamics.

\subsection{Origin of equivalence between continuous and
pulsed formulations}
\label{sec-unitequiv}

The equivalence between the pulsed and continuous measurement
formulation of the quantum Zeno effect can be pushed much further:
let us show that the two procedures differ only in the order in
which two limits are computed \cite{FLP}. As we have seen, the
continuous case deals with the strong coupling limit
\begin{equation}
H_K=H+KH_{\mathrm{c}},\qquad K\rightarrow \infty
\label{eq:cont}
\end{equation}
and the Zeno subspaces are the eigenspaces of $H_{\mathrm{c}}$. On
the other hand, the kicked dynamics entails the limit
$N\rightarrow
\infty $ in (\ref{eq:BBevol}) and the Zeno subspaces are the
eigenspaces of $U_{\mathrm{kick}}$. This evolution is generated by
the Hamiltonian
\begin{equation}
H_{\mathrm{kick}}=H+\tau_{0}H_{\mathrm{c}}\sum_{n}\delta (t-n\tau
),\qquad \tau \rightarrow 0  \label{eq:puls}
\end{equation}
where $\tau$ is the period between two kicks and the unitary
evolution during a kick is $U_{\mathrm{kick}}=\exp
(-i\tau_0 H_{\mathrm{kick}})$. The limit $N\rightarrow \infty $
in (\ref{eq:BBevol}) corresponds to $\tau\rightarrow 0$. The
two dynamics (\ref{eq:cont}) and (\ref{eq:puls}) are both limiting
cases of the following one
\begin{eqnarray}
  H(\tau,K) &=& H+KH_{\mathrm{kick}}
  \sum_{n}g\left( \frac{t-n(\tau+\tau_0/K)}{\tau_{0}/K}\right) ,
\label{eq:contpuls}
\end{eqnarray}
where the function $g$ has the properties
\begin{eqnarray}
\sum_{n}g(x-n) &=& 1 \label{eq:prop1}\\
\lim_{K\rightarrow \infty} K g(K x) &=& \delta (x).
  \label{eq:prop2}
\end{eqnarray}
For example we can consider $g(x)=\chi_{[-1/2,1/2]}(x)$. In Eq.\
(\ref{eq:contpuls}) the period between two kicks is $\tau_{0}/K+\tau$,
while the kick lasts for a time $\tau_{0}/K$.
By taking the limit $\tau\rightarrow 0$ in Eq.\
(\ref{eq:contpuls}), i.e., a sequence of pulses of finite duration
$\tau_0/K$ without any idle time among them, and using property
(\ref{eq:prop1}), one recovers the continuous case
(\ref{eq:cont}). Then, by taking the strong coupling limit
$K\rightarrow \infty $ one gets the Zeno subspaces. On the other
hand, by taking the $K\rightarrow \infty $ limit, i.e., the limit
of shorter pulses (but with the same global---integral---effect),
and using property (\ref{eq:prop2}) and the identity $\delta
(t/\tau_0)=\tau_{0}\delta (t)$, one obtains the kicked case
(\ref{eq:puls}). Then, by taking the vanishing idle time limit
$\tau\rightarrow 0$ one gets again the Zeno subspaces. In
short, the mathematical equivalence between the two approaches is
expressed by the relation
\begin{equation}
\label{eq:limits}
  \lim_{K\rightarrow \infty }\;\lim_{\tau \rightarrow
    0}H(\tau,K)
  =\lim_{\tau\rightarrow 0}\;\lim_{K\rightarrow \infty }H(\tau,K),
\end{equation}
(for almost all $\tau_0$) with the left (right) side expressing the
continuous (pulsed) case. Note that this formal equivalence must
physically be checked on a case by case basis, and it is legitimate
only if the inverse Zeno regime is avoided and the role of the form
factors clearly spelled out. That is, physically the relevant
timescales play a crucial role, and in practice there certainly can
be a difference \cite{FTPNTL2005} between kicked dynamics and
continuous coupling, in spite of their equivalence in the above
mathematical limit.

\section{Examples}
 \label{sec-example}
 \andy{example}

One of the main potential applications of the quantum Zeno
subspaces\index{quantum Zeno subspaces} concerns the possibility
of freezing the loss of quantum mechanical coherence and
probability leakage due to the interaction of the system of
interest with its environment. Let us therefore look at some
elementary examples in the light of the three different
formulations of the Zeno effect summarized in Sec.\
\ref{sec-zenosubsp}. In the following, it can be
helpful to think of the Zeno subspace $\cH_1$ as the quantum
computation subspace (qubit) that one wants to protect from
decoherence.

\subsection{Von Neumann's projections}
\label{sec-examplevn}
\andy{examplevn}

\begin{figure}[t]
\begin{center}
\includegraphics[height=3cm]{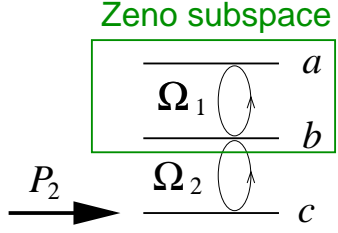}
\end{center}
\caption{Three level system undergoing measurements ($P_1$ not
indicated). We explicitly showed the Zeno subspace $\cH_1$.}
\label{fig:fig2}
\end{figure}

Consider a 3-level system in $\cH_{\mathrm{sys}}=\mathbb{C}^3$
\andy{3level}
\beq
\bra{a} = (1,0,0), \quad \bra{b} = (0,1,0), \quad \bra{c} =
(0,0,1)
\label{eq:3level}
\eeq
and the  Hamiltonian
\andy{ham3lev}
\beq
H = \Omega_1 ( \ket{a} \bra{b} + \ket{b} \bra{a}) + \Omega_2 (
\ket{b} \bra{c} + \ket{c} \bra{b}) = \pmatrix{0 & \Omega_1 & 0 \cr
\Omega_1 & 0 & \Omega_2 \cr 0 & \Omega_2 & 0 }.
\label{eq:ham3lev}
\eeq
We perform the (incomplete, nonselective) projective measurements
($P_1+P_2 =\mathbf{1}$)
\andy{proj3}
\beq
P_1 = \ket{a} \bra{a} + \ket{b} \bra{b} = \pmatrix{1 & 0 & 0 \cr 0 &
1 & 0 \cr 0 & 0 & 0 }, \quad P_2 = \ket{c} \bra{c}= \pmatrix{0 & 0 &
0 \cr 0 & 0 & 0 \cr 0 & 0 & 1 },
\label{eq:proj3}
\eeq
yielding the partition (\ref{eq:partition}), with $\mathrm{dim}
\cH_{1}=2$, $\mathrm{dim} \cH_{2}= 1$. The evolution operators
(\ref{eq:cVfin}) read
\andy{evolmeas3}
\barr
U_{\rm Z}^{(1)}(t) &= P_1 \exp(-i P_1 H P_1 t) &= P_1\exp
\left[-i\pmatrix{0 &
\Omega_1t & 0 \cr \Omega_1t & 0 & 0 \cr 0 & 0 & 0 } \right]
\nonumber\\
&  &=\pmatrix{\cos\Omega_1t & -i\sin\Omega_1t & 0 \cr
-i\sin\Omega_1t &
\cos\Omega_1t & 0 \cr 0 & 0 & 0 }, \nonumber \\
U_{\rm Z}^{(2)}(t) &= P_2 \exp(-i P_2 H P_2 t) &= P_2= \pmatrix{0 & 0
& 0 \cr 0 & 0 & 0 \cr 0 & 0 & 1 }
\label{eq:evolmeas3}
\earr
and the Zeno Hamiltonian (\ref{eq:ZenoHam}) is
\beq
H_{\mathrm{Z}}= P_1 H P_1 + P_2 H P_2 =\pmatrix{0 & \Omega_1 & 0
\cr \Omega_1 & 0 & 0 \cr 0 & 0 & 0 }.
\eeq
The initial state (\ref{eq:inrho}) evolves according to
(\ref{eq:rhoZ}): in the Zeno limit ($N\to \infty$), the subspaces
$\cH_{1}$ and $\cH_{2}$ decouple. If the coupling $\Omega_2$ is
viewed as a caricature of the loss of quantum mechanical
coherence, the subspace $\cH_{1}$ becomes ``decoherence free"
\cite{palma,zanardi,duan}. See Fig.\ \ref{fig:fig2}.

\subsection{Kicks}
\label{sec-examplek}
\andy{examplek}

In order to exemplify how unitary kicks yield the Zeno subspaces,
consider the 4-level system in the enlarged Hilbert space
$\cH_{\mathrm{sys}}\oplus \mathrm{span}\{\ket{M}\}$
\andy{levelM}
\barr
& & \bra{a} = (1,0,0,0), \quad \bra{b} = (0,1,0,0), \nonumber \\
& & \bra{c} = (0,0,1,0), \quad \bra{M} = (0,0,0,1)
\label{eq:levelM}
\earr
and the Hamiltonian
\andy{ham3l0}
\beq
H = \Omega_1 ( \ket{a} \bra{b} + \ket{b} \bra{a}) + \Omega_2 (
\ket{b} \bra{c} + \ket{c} \bra{b}) = \pmatrix{0 & \Omega_1 & 0 &
0\cr \Omega_1 & 0 & \Omega_2 & 0 \cr 0 & \Omega_2 & 0 & 0 \cr 0 & 0
& 0 & 0 } .
\label{eq:ham3l0}
\eeq
This is the same example as (\ref{eq:3level})-(\ref{eq:ham3lev}),
but we added a fourth level $\ket{M}$. We now couple $\ket{M}$ to
$\ket{c}$ by performing the unitary kicks
\andy{eq:3kicks}
\barr
U_{\mathrm{kick}}&=&  P_1 + \mathrm{e}^{-i \lambda (\ket{c}
\bra{M} + \ket{M} \bra{c})} = \pmatrix{1 & 0 & 0 &
0\cr 0 & 1 & 0 & 0 \cr 0 & 0 & \cos\lambda & -i\sin\lambda \cr 0 &
0 & -i\sin\lambda & \cos\lambda }
\nonumber\\
&=& \sum_{n=1,\pm} \mathrm{e}^{-i\lambda_n} P_n,
\label{eq:3kicks}
\earr
where $\lambda=\lambda_+ = -\lambda_- \neq
\lambda_1=0$ and the subspaces are defined by
\andy{eq:meassub}
\begin{eqnarray}
 P_1 &=& \ket{a} \bra{a} + \ket{b} \bra{b}= \pmatrix{1 & 0
& 0 & 0\cr 0 & 1 & 0 & 0 \cr 0 & 0 & 0 & 0 \cr 0 & 0 & 0 & 0 } ,
\label{eq:meassuba} \\
 P_\pm &=& \frac{(\ket{c} \pm \ket{M})(\bra{c} \pm
\bra{M})}{2} = \frac{1}{2}\pmatrix{0 & 0 & 0 & 0\cr 0 & 0 & 0 & 0
\cr 0 & 0 & 1 & \pm 1 \cr 0 & 0 & \pm 1 & 1 } .
\label{eq:meassubb}
\end{eqnarray}
($P_1 + P_- + P_+=\mathbf{1}$.)

\begin{figure}[t]
\begin{center}
\includegraphics[height=3cm]{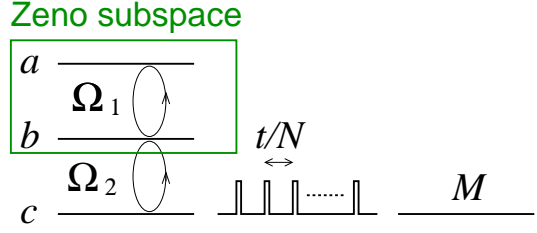}
\end{center}
\caption{Three level system undergoing frequent unitary kicks that
couple one of its levels to an ``external" system $M$. We
explicitly indicated the Zeno subspace $\cH_1$.}
\label{fig:fig3}
\end{figure}

In the Zeno limit ($N\to \infty$) the subspaces $\cH_1$, $\cH_{+}$
and $\cH_{-}$ decouple due to the wildly oscillating phases
O$(N)$. See Fig.\ \ref{fig:fig3}. The Zeno Hamiltonian
(\ref{eq:eqHz}) reads
\andy{eq:eqHz}
\beq
\label{eq:eqHzpp}
H_{\mathrm{Z}} = \sum_n P_n H P_n = \pmatrix{0 & \Omega_1 & 0 & 0
\cr \Omega_1 & 0 & 0 & 0 \cr 0 & 0 & 0 & 0 \cr 0 & 0 & 0 & 0 }
\eeq
and the evolution (\ref{eq:evolNZ}) is
\andy{eq:evolexNZ}
\barr
U_N(t) & \sim & \exp\left(-i \sum_n  N \lambda_n P_n + P_n H P_n t
\right) \nonumber \\
& = & \exp \left[-i\pmatrix{0 & \Omega_1t & 0 & 0\cr
\Omega_1t & 0 & 0 & 0 \cr 0 & 0 & 0 & N\lambda
\cr 0 & 0 & N\lambda &
0 } \right]  \nonumber\\
&=&\pmatrix{\cos\Omega_1t & -i\sin\Omega_1t & 0 & 0\cr
-i\sin\Omega_1t & \cos\Omega_1t & 0 & 0 \cr 0 & 0 & \cos N
\lambda & -i\sin N \lambda \cr 0 & 0 & -i\sin N \lambda & \cos N \lambda} .
\label{eq:evolexNZpart}
\earr
This is the scheme adopted by Itano \emph{et al} in their
experiment \cite{Itano90}.

\begin{figure}[t]
\begin{center}
\includegraphics[height=3cm]{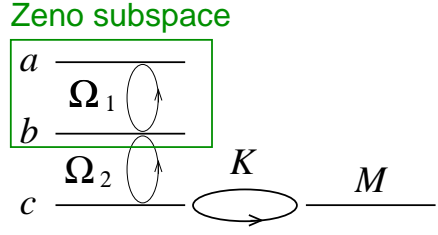}
\end{center}
\caption{Three level system with one of its levels strongly
coupled to an ``external" system $M$.  We explicitly indicated the
Zeno subspace $\cH_1$.}
\label{fig:fig4}
\end{figure}

\subsection{Continuous coupling}
\label{sec-examplec}
\andy{examplec}

Finally, in order to understand how the scheme involving
continuous measurements works, add to (\ref{eq:ham3l0}) the
Hamiltonian (acting on $\cH_{\mathrm{sys}}\oplus
\mathrm{span}\{\ket{M}\}$)
\andy{hmeas}
\beq
KH_{\mathrm{c}} = K( \ket{c} \bra{M} + \ket{M} \bra{c}) =
\pmatrix{0 & 0 & 0 & 0\cr 0 & 0 & 0 & 0 \cr 0 & 0 & 0 & K \cr 0 &
0 & K & 0 } = K\left(P_{+}-P_{-}\right) \; ,
\label{eq:hmeas}
\eeq
where $P_\pm$ are the same as in (\ref{eq:meassubb}). The fourth
level $\ket{M}$ is now continuously coupled to level $\ket{c}$,
$K \in \mathbb{R}$ being the strength of the coupling. As $K$ is
increased, level $\ket{M}$ performs a better ``continuous
observation" of $\ket{c}$, yielding the Zeno subspaces. The
eigenprojections of $H_{\mathrm{c}}$ [see (\ref{eq:diagevol})]
\beq
H_{\mathrm{c}}= \eta_1 P_1+ \eta_{-} P_{-} + \eta_{+} P_{+}
\eeq
are again (\ref{eq:meassuba})-(\ref{eq:meassubb}), with $\eta_1=0,
\eta_\pm = \pm 1$. Once again, in the Zeno limit ($K\to \infty$)
the subspaces $\cH_{1}$, $\cH_{+}$ and $\cH_{-}$ decouple due to the
wildly oscillating phases O$(K)$. See Fig.\ \ref{fig:fig4}. The Zeno
Hamiltonian $H_{\mathrm{Z}}$ is given by (\ref{eq:diagsys}) and
turns out to be identical to (\ref{eq:eqHzpp}), while the evolution
(\ref{eq:measinterbis}) explicitly reads
\andy{eq:evolcontZ}
\barr
U_K(t) &\sim & \exp\left(-i \sum_n  Kt \eta_n P_n + P_n H P_n t
\right) \nonumber \\
& = & \exp \left[-i\pmatrix{0 & \Omega_1t & 0 & 0\cr
\Omega_1t & 0 & 0 & 0 \cr 0 & 0 & 0 & Kt \cr 0 & 0 & Kt &
0 } \right]
\nonumber\\
&=&\pmatrix{\cos\Omega_1t & -i\sin\Omega_1t & 0 & 0\cr
-i\sin\Omega_1t & \cos\Omega_1t & 0 & 0 \cr 0 & 0 & \cos Kt & -i\sin
Kt \cr 0 & 0 & -i\sin Kt & \cos Kt }.
\label{eq:evolcontZ}
\earr
[Compare with (\ref{eq:evolexNZpart}): $Kt$ plays the role of $N
\lambda$.] This is the scheme adopted by Ketterle and collaborators in their
experiment \cite{ketterle}.

\section{Conclusions and outlook}
\label{sec-final}

We analyzed the physical and mathematical aspects of the quantum
Zeno dynamics that takes place when one frequently checks whether a
quantum system has remained inside a multidimensional Zeno
subspace. Unlike in the traditional formulation of the QZE, the
system can evolve away from its initial state, although it remains
in the eigenspace of the projection operator associated with the
measurement.

When the Zeno subspace is finite dimensional, the evolution can be
easily (and rigorously) derived. The situation is much more
complicated for infinite dimensional projections, such as
traditional position measurements, namely projections onto spatial
regions. This is an open problem from the mathematical point of
view, where the existence of the strong limit of the Zeno product
formula remains to be proven. However, if it converges, the Zeno
dynamics uniquely determines the boundary conditions, and they turn
out to be of Dirichlet type.

The Zeno mechanism not only forces the system to remain in a given
subspace, it also constrains its (sub)dynamics in this space,
determining the behavior of the wave function on the boundary and
yielding a unitary, decoherence free evolution. Besides its
theoretical interest, this feature might lead to potential
applications and practical implementations of the Zeno constraints
in order to tailor subspaces that are robust against decoherence,
which are of great interest in quantum information processing
applications.

We implicitly assumed, throughout a part of our discussion, the
validity of the Copenhagen interpretation, according to which the
measurement is considered to be instantaneous. The QZE is
traditionally derived by considering a series of rapid, pulsed
observations (projections). This became almost a dogma and
motivated all seminal experiments. However, a projection operator is
a shorthand notation, that summarizes the effects of a much more
complicated underlying dynamical process, involving a huge number of
elementary quantum mechanical systems. Later formulations emphasized
that the QZE can also be generated by pulsed and even continuous
Hamiltonian interaction. Here we have shown that all these seemingly
different pictures can be unified and in particular the QZE in its
continuous-interaction and pulsed (``kicks'' or ``bang-bang'')
formulation can be understood as limits of a single Hamiltonian,
Eq.~(\ref{eq:contpuls}), giving rise to either pulsed or continuous
dynamics, with a resulting partitioning of the controlled system's
Hilbert space into quantum Zeno subspaces. This unified view not
only offers the advantage of conceptual simplicity, but also has
significant practical consequences: it shows that the scope of all
the methods analyzed here (QZE, kicks and continuous interaction)
are wider than previously suspected, leading to greater flexibility
in their implementation.

The present work enters an experimentally uncharted area, although
the property of being a multidimensional measurement is not at all
exotic: the quantum Zeno dynamics has not been experimentally
demonstrated, even for a two-dimensional subspace (a qubit). It
would be of great interest to verify it for an $N$ level system or
for a collection of qubits and in particular for the most basic
quantum measurement: position.

\ack
We would like to thank G. Badurek, R. Fazio, G. Florio, Z. Hradil, D. A. Lidar,
G. Marmo, H. Nakazato, M. Namiki, I. Ohba, J. Pe\v{r}ina, H. Rauch,
J. \v{R}eh\'{a}\v{c}ek, A. Scardicchio, L. S. Schulman, E.G.C.
Sudarshan, S. Tasaki and K. Yuasa for many conversations on the Zeno
phenomenon. Figure \ref{fig:atene} is reproduced by courtesy of the
Vatican Museums. This work is partly supported by the European
Community through the Integrated Project EuroSQIP, by the bilateral
Italian Japanese Projects II04C1AF4E on ``Quantum Information,
Computation and Communication" of the Italian Ministry of
Instruction, University and Research, and the Joint Italian Japanese
Laboratory on ``Quantum Information and Computation" of the Italian
Ministry for Foreign Affairs.

\end{document}